\documentclass[11pt]{article}

\usepackage[utf8]{inputenc}
\usepackage{fullpage}
\usepackage{titling} 
\usepackage{graphicx}
\usepackage{siunitx} 
\usepackage{rotating} 

\usepackage{natbib}
\bibpunct{(}{)}{,}{a}{}{;}

\usepackage[page]{appendix}
\usepackage{amsmath}
\usepackage{amssymb}
\usepackage{amsthm}
\usepackage[labelfont=bf]{caption}
\usepackage{booktabs}
\usepackage{threeparttable}

\newtheorem{prop}{Proposition}

\usepackage[hidelinks]{hyperref}

\title{Latent Stratification for Incrementality Experiments\thanks{We appreciate suggestions we received from seminar and conference participants at Drexel University, University of Chicago, Imperial College London, Ohio State University, Northwestern University, University of Rochester, University of Southern California, Southern Methodist University, Stanford University, Tel-Aviv University, Temple University, University of Warwick, MIT CODE, the Virtual Quant Marketing Seminar, Marketing Science and the Consumer Analytics Workshop in Chile. Hangcheng Zhao and Bolun Xiao provided excellent research support. Christophe Van den Bulte provided valuable feedback. Wharton Customer Analytics provided assistance in obtaining data. Funding was provided by an Adobe Digital Experience Research Award and a Dean's Fellowship from LeBow College of Business.}}

\author{
Ron Berman \\ 
The Wharton School \\ 
University of Pennsylvania \\ 
ronber@wharton.upenn.edu 
\and 
Elea McDonnell Feit \\
LeBow College of Business \\ 
Drexel University \\ 
eleafeit@gmail.com 
}

\date{May 2023}

\begin{document}
\maketitle
\thispagestyle{empty}

\newpage
\setcounter{page}{1}
\begin{center}
\Large
\thetitle
\end{center}
\vspace{0.5in}
\begin{abstract}
Incrementality experiments compare customers exposed to a marketing action designed to increase sales to those randomly assigned to a control group. These experiments suffer from noisy responses which make precise estimation of the average treatment effect (ATE) and marketing ROI difficult. We develop a model that improves the precision by estimating separate treatment effects for three latent strata defined by potential outcomes in the experiment -- customers who would buy regardless of ad exposure, those who would buy only if exposed to ads and those who would not buy regardless. The overall ATE is estimated by averaging the strata-level effects, and this produces a more precise estimator of the ATE over a wide range of conditions typical of marketing experiments. Analytical results and simulations show that the method decreases the sampling variance of the ATE most when (1) there are large differences in the treatment effect between latent strata and (2) the model used to estimate the strata-level effects is well-identified. Applying the procedure to 5 catalog experiments shows a reduction of 30-60\% in the variance of the overall ATE. This leads to a substantial decrease in decision errors when the estimator is used to determine whether ads should be continued or discontinued.
\end{abstract}
\textbf{Keywords:} Advertising, incrementality experiments, lift testing, A/B testing, holdout experiments, average treatment effect, principal stratification, causal inference.

\newpage
\section{Introduction}

Incrementality experiments (also called lift tests or holdout experiments) gauge the effect of a marketing action (e.g. direct marketing, exposure to an ad, a discount offer) by comparing sales for customers who are randomly assigned to receive the marketing communication to sales for customers who do not receive that communication, i.e. a control or holdout group \citep{LewisRao2015, SahniZouChintagunta2016, JohnsonReileyNubbmeyer2017, JohnsonLewisReiley2017, LewisWong2018}. The estimated sales lift obtained from such experiments can be compared to advertising costs to determine whether the advertising has positive returns. This approach to gauging the value of advertising is the ``gold standard" for causal inference \citep{gordonetal2019}, is increasingly popular among digital marketers \citep{AdExchanger2021}, and has been implemented in tools like Facebook Conversion Lift and Google Campaign Experiments.

We focus on experiments where the treatment is randomized at the customer level and the response variable is customer-level sales or another continuous, non-negative variable for each customer such as time-on-site. Such experiments hold a lot of promise for helping firms decide which marketing communications are worthwhile. However, in practice, estimates of sales lift from these experiments often have high sampling variation because the response of individual customers has high variance and advertising effects are small relative to this noise. Even with very large experiments, this can result in estimates of ROI that are not statistically significant \citep{LewisRao2015}. Compounding this problem, marketers often use small control groups (e.g. 5-10\% of the total test size) to reduce the opportunity cost of the test \citep{FeitBerman2019}. As we show, this leads to more errors and reduced profit when such estimates are used for decision making. 

A consistent feature of customer-level sales is the large number of observed zeros; many customers in the target audience for an incrementality test do not make a purchase. In the catalog holdout experiments we analyze, more than 80\% of customers do not make a purchase, even though all of them have purchased from the retailer previously. Many reported marketing experiments have similarly low levels of purchases \citep[e.g.][]{HobanBucklin2015, ZantedeschiFeitBradlow2016, SahniZouChintagunta2016, JohnsonReileyNubbmeyer2017}. We exploit this feature and propose a new model for analyzing incrementality experiments that divides customers into three latent strata based on their unobserved potential outcomes:\footnote{The Neyman-Rubin potential outcomes represent the customer's response under the assigned and counterfactual treatments in the focal experiment \citep[cf.][Section 1.3]{ImbensRubin2015}. Prior to randomization, the potential outcomes for a given experiment are unknown but fixed (not random variables). After randomization, the potential outcome corresponding to the assigned treatment is observed for each customer.} ``$A$'' customers would make a purchase in the experiment regardless of assignment to treatment or control, but perhaps at different amounts, ``$B$'' customers would only purchase if they are assigned to treatment, and ``$C$'' customers would not purchase in the experiment regardless of treatment.

Dividing customers into strata based on potential outcomes is called principal stratification \citep{FrangakisRubin2002}. Readers may be familiar with principal stratification as an interpretation of IV estimation for non-compliance. In that setting, customers are divided into principal strata based on treatment compliance. Here, we stratify based on \emph{whether or not the outcome is positive}.\footnote{Appendix \ref{sec:principal_stratification} summarizes differences between latent stratification and other applications of principal stratification.} While customers' strata membership is not directly observed, the size and treatment effect for each stratum can be estimated using a parametric model. Understanding the size and average treatment effect for strata $A$ and $B$ provides advertisers insight into whether their ads encourage customers to buy or encourage those already buying to buy more. That is, the analysis tells us how much ads operate on the extensive margin versus the intensive margin. 

Importantly, this novel application of principal stratification allows advertisers to estimate the overall average treatment effect (ATE) more precisely, by taking a weighted average of the treatment effects for the latent strata. The strata have radically different treatment effects: customers in stratum $B$ buy only under treatment and have a relatively large average treatment effect, while customers in stratum $C$ all have a treatment effect of exactly zero. Thus the individual treatment effect for the strata are more homogeneous within each stratum and heterogeneous between strata, which is precisely the conditions under which post-stratification\footnote{In a post-stratified analysis, customers are divided into strata based on \emph{observed} pre-randomization covariates and the overall treatment effect is computed as the weighted average of the groups. For applications of post-stratification in marketing see \cite{simester2020efficiently} and \cite{gordon2022close}.} produces a more precise estimate of the ATE \citep{MiratrixSekhonYu2013}. By taking a weighted average of the treatment effects for each stratum, we estimate the overall ATE more precisely than the usual difference-in-means estimator. The mechanism for the improved precision of the ATE from latent stratification is similar to that of stratification on observed pre-randomization covariates, e.g. CUPED \citep{deng2013}. However, latent stratification accomplishes this without the need for tracking user covariates, which is not only costly, but can raise data storage and privacy issues. Further, stratifying on observed pre-randomization covariates may actually increase sampling variation if the covariates are not correlated with the outcome, unless a sophisticated variable selection method is used \citep{guo2021machine}. As we show in Section \ref{sec:precision}, post-stratification on the latent strata would \emph{always} decrease the variance of the ATE, if the latent strata were observed.

Of course, we do not observe users' potential outcomes or latent strata, but we can estimate the size and the treatment effect for each stratum by making several additional assumptions which are laid out in Section \ref{sec:model}. Specifically, we assume that the positive response under treatment follows a parametric distribution which is different for strata $A$ and $B$. The fact that these distributions are estimated without observing the strata membership for individuals introduces some additional uncertainty that could countervail the improvement in the precision of ATE that would be achieved if the latent strata were observed. To estimate the real-world performance of latent stratification, we test it empirically with a number of simulated data sets designed to represent typical marketing incrementality experiments. The method nearly always improves the precision of the ATE across the scenarios we tested. The simulation study reported in Section \ref{sec:precision} shows that the greatest improvements in precision occur when the distributions of outcomes for $A$ and $B$ are more distinct. 

Latent stratification is carried out in the analysis stage of the experiment, and requires no changes to the design of the experiment; the strategies we propose can be used to re-analyze experiments that have already been fielded resulting in an estimate with lower sampling variation. The main downside is that latent stratification may result in a biased estimate of the ATE if the model is misspecified.\footnote{The estimator is asymptotically unbiased (consistent) if the model is correctly specified.} However, we mitigate this risk by providing a misspecification test. 

We use the model to analyze 5 catalog incrementality experiments and show that the latent stratification model improves the variance of the estimated ATE by 30-60\% relative to the difference-in-means estimator. This isn't a statistical nicety; as we illustrate, this more precise estimate leads to better decisions about whether to discontinue advertising. In a representative simulation that draws from the real data, we show that the error rate can be reduced from 7.5\% to 0.5\%. This level of variance reduction is not achieved by sophisticated methods that use covariates to reduce the variance of the ATE. Nor is it achieved by a model that posits separate ``spike-at-zero" distributions for the treatment and control groups, to account for zero outcomes without considering the latent strata.  

In the next section, we lay out the latent stratification model in detail along with a misspecification test to ensure that the model is appropriate for the data. In the following section, we apply the method to 5 catalog incrementality tests, and compare several benchmarks, some of which use covariates, to latent stratification. In the following section, we explain when the latent stratified estimate of the ATE provides a reduction in variance over the standard difference-in-means estimate of the ATE. In the final section, we review the procedure step-by-step and discuss the limitations and potential extensions of the approach. 

\section{Latent stratification}
\label{sec:model}
\subsection{The marketer's decision problem}  

Consider a marketer who is deciding whether to continue using a particular marketing channel or to discontinue using it for all customers. To decide which action to take, the marketer runs an incrementality experiment on $n$ customers indexed by $i = 1,\ldots,n$ that are randomly assigned to the treatment which exposes them to the marketing channel ($Z_i = 1$) or to the control which holds them out from exposure ($Z_i = 0$). A non-negative outcome $Y_i(Z_i)$ is observed for each customer. We assume that the marketer can determine the duration over which to measure the response to the treatment based on previous analysis \citep[e.g.][]{BonfrerDreze2009, ZantedeschiFeitBradlow2016} or has a surrogate model for longer-range outcomes \citep{Atheyetal2016, yang2020targeting}.

The primary goal of the experiment is to estimate the average treatment effect (ATE) for the targeted population: 
\begin{equation}
\text{ATE} \equiv \tau \equiv \mathbb{E}[Y_i(Z_i=1)-Y_i(Z_i=0)]
\label{eq:ATE_defn}
\end{equation}
where $Y_i(Z_i=0)$ and $Y_i(Z_i=1)$ are the unobserved potential outcomes for customer $i$. The ATE is important because it can be compared to the costs of the marketing action to determine if the treatment produces enough incremental profit to exceed the costs of the marketing. If the treatment provides enough incremental profit, it is optimal to deploy the treatment going forward, and otherwise it is better to cease using the treatment, i.e. deploy the control.

The ATE is usually estimated by the difference-in-means (DiM) estimator: 
\begin{equation}
\label{eq:diff-in-means_est}
\widehat{\tau}^{\text{DiM}} \equiv \frac{1}{n_1} \sum_{i|Z_i=1} Y_i - \frac{1}{n_0} \sum_{i|Z_i=0}  Y_i
\end{equation}
where $n_1$ and $n_0$ are the numbers of customers assigned to treatment and control and $Y_i$ is the observed outcome for customer $i$ under random assignment to treatment. This estimator can also be obtained by linear regression of the form $Y_i = \widehat{\alpha} +\widehat{\tau}^{\text{DiM}} Z_i + \epsilon_i$. 

Under the assumption that the outcome for each customer depends only on their treatment assignment $Z_i$ and not on the assignment for other customers (i.e. SUTVA\footnote{Stable Unit Treatment Value Assumption.}), this estimator is unbiased. The marketer could use the value of $\widehat{\tau}^{\text{DiM}}$ as a point estimate of the ATE in their decision making and deploy the treatment if $\widehat{\tau}^{\text{DiM}}$ exceeds the cost of treatment.

The standard estimator for the sampling variance of $\widehat{\tau}^\text{DiM}$ is:
\begin{equation}
\label{eq:diff-in-means_var}
Var(\widehat{\tau}^{\text{DiM}}) \equiv \frac{s_1^2}{n_1} + \frac{s_0^2}{n_0}
\end{equation}
where $s_1$ and $s_0$ are the standard deviations of the response in each group. This is a conservative estimate of the variance \citep[cf.][Chapter 6]{ImbensRubin2015}.

A longstanding problem in incrementality experiments is that the sampling variance of $\widehat{\tau}^{\text{DiM}}$ is often large due to the large variance of $Y_i$ in most marketing response data \citep{LewisRao2015, Yang2021}. The implication is that although $\widehat{\tau}^{\text{DiM}}$ is unbiased, because of its large sampling variance, there is a high probability that $\widehat{\tau}^{\text{DiM}}$ might be small when the true ATE is large, and vice versa, causing the marketer to take the wrong action. If an alternative estimator has lower sampling variance, then it is more likely that the marketer will pick the action that matches the true optimal action, and hence achieve higher profits.
In the next subsection we propose an alternative estimator, the latent stratified estimator, which has lower sampling variation than $\widehat{\tau}^{\text{DiM}}$. 

\subsection{Latent stratified (LS) estimator of the ATE}
\label{sec:ls}
We focus on experiments where the outcome $Y_i$ is a positive, continuous value like sales to the customer, visit time on a website or amount of content consumed. Continuous outcomes like sales are often used in incrementality experiments. Unlike binary measures such as a purchase or a click, continuous outcomes are more informative and have a direct economic meaning that can be compared to costs. 

\begin{figure}[htbp]
\centering
\includegraphics{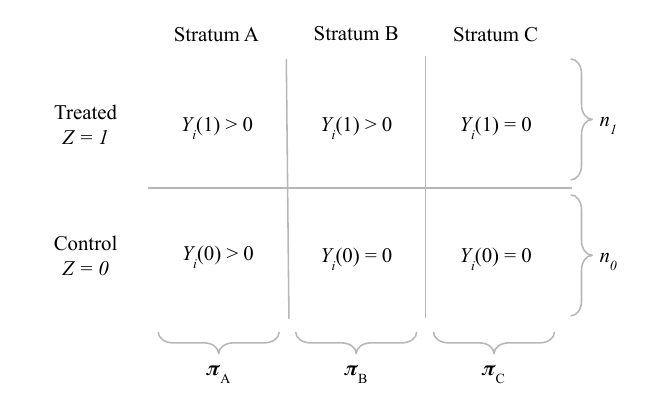}
\caption{Latent strata}
\label{fig:latent_strata}
\end{figure}

For an outcome like sales, it is often observed that many customers do not purchase, resulting in many observed zeros for $Y_i$. For example, in the experiments we analyze in Section \ref{sec:applications}, the response is customer purchases and more than 80\% of customers do not purchase, resulting in an outcome of $Y_i=0$. Similarly, when the outcome is time on site or content consumed, a large fraction of outcomes will be zero indicating that the customer did not engage in the activity. Latent stratification leverages this fact, conceptually dividing customers into three strata ($A$, $B$ and $C$) based on whether customers' potential outcomes are zero or positive:\footnote{Potential outcomes are not random variables, but fixed quantities that are unobserved prior to randomization. Stratifying on them does not result in bias \citep{FrangakisRubin2002}.} 
\begin{equation}
\begin{aligned}
i \in A \iff Y_i(1)>0 \And Y_i(0)>0 \\
i \in B \iff Y_i(1)>0 \And Y_i(0)=0\\
i \in C \iff Y_i(1)=0 \And Y_i(0)=0
\end{aligned}
\end{equation}
If the outcome is purchase amount, this means that customers in $A$ purchase regardless of treatment, customers in $B$ purchase only if treated and customers in $C$ do not purchase regardless of treatment.\footnote{The potential outcomes are defined with respect to a single experiment and are not a persistent property of a customer. A customer might be in stratum $A$ for an experiment conducted during the peak sales period and in stratum $C$ for another experiment outside the peak period.} Each customer belongs to one latent stratum; the proportion of customers in each stratum is $\pi_A$, $\pi_B$ and $\pi_C$ (see Figure \ref{fig:latent_strata}). 

Our goal is to compute a stratified estimate of the ATE:  
\begin{equation}
\widehat{\tau}^{\text{LS}} \equiv \widehat{\pi}_A \widehat{\tau}_A + \widehat{\pi}_B \widehat{\tau}_B + \widehat{\pi}_C \widehat{\tau}_C 
\label{eq:stratified_ATE}
\end{equation}
where $\widehat{\tau}_A$, $\widehat{\tau}_B$ and $\widehat{\tau}_C$ are estimates of the ATE for each stratum and $\widehat{\pi}_{A}$, $\widehat{\pi}_{B}$ and $\widehat{\pi}_{C}$ are estimates of the strata proportions.  

As we show in Section \ref{sec:precision}, under the hypothetical scenario where the stratum for each customer is observed, the stratified estimator will produce a more precise estimate of the ATE. Intuitively, the estimate is more precise because the individual treatment effects are more homogeneous within-strata and heterogeneous between-strata, which is precisely the conditions where a stratified estimator will have lower variance than the DiM estimator \citep{MiratrixSekhonYu2013}. To see this, notice that by definition, all potential outcomes in $C$ are zero, which implies that the individual treatment effects in $C$ have a mean and variance of zero. In contrast, the individual treatment effects in $B$ are $Y_i(1) - 0$, which will typically have a mean that is quite far from zero because people who purchase tend to purchase substantial amounts. And in $A$, the individual treatment effects are $Y_i(1) - Y_i(0)$, and their average will typically be smaller than in $B$. So, by construction, we have created strata where the variance of the individual treatment effects are smaller within-strata and larger across-strata. 

\begin{figure}[htbp]
\centering
\includegraphics{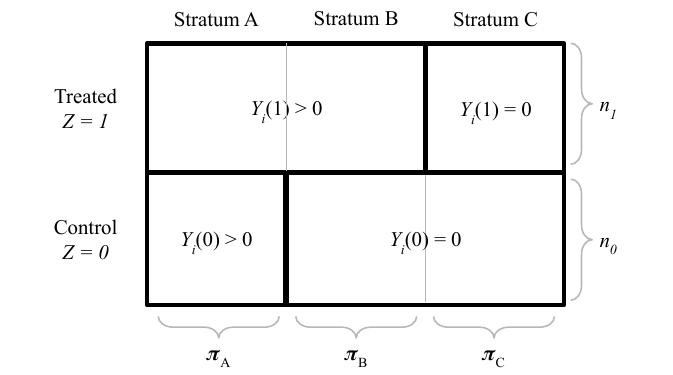}
\caption{Latent strata and observational groups. Dark rectangles denote the four groups that can be observed for an experiment. }
\label{fig:obs_groups}
\end{figure}

In reality, we can never observe both potential outcomes and so we never directly observe the stratum membership of each customer. However, based on the observed treatment assignment and outcomes, we can divide the customers into the four observational groups outlined in Figure \ref{fig:obs_groups}. 
The strata proportions $\pi_A$, $\pi_B$ and $\pi_C$ can be estimated directly from the the sizes of these observational groups, as follows. Treated customers who do not make a purchase (upper right in Figure \ref{fig:obs_groups}) must belong to $C$, and therefore we can estimate $\pi_C$ by the proportion of treated customers who do not make a purchase. The proportion $\pi_A$ can be estimated similarly, as the proportion of control customers that make a purchase (lower left in Figure \ref{fig:obs_groups}). An estimate of $\pi_B$ can then be computed as $1-\pi_A-\pi_C$. 

The challenge in latent stratification is estimating $\tau_A$ and $\tau_B$ (while $\tau_C$ is zero by definition). The treated customers who purchase (upper left in Figure \ref{fig:obs_groups}) can belong to $A$ or $B$ and so it is impossible to estimate $\tau_A$ and $\tau_B$ without further assumptions. One solution is to construct bounds on $\tau_A$ and $\tau_B$ by assuming that the lowest (highest) $\pi_B$ observed positive purchase amounts belong to $B$, while the remaining belong to $A$. This produces an upper (lower) bound on the mean of the outcomes for each group, which in turn can be used to construct bounds on $\tau_A$, $\tau_B$ and $\tau^{\text{LS}}$. However, these non-parametric bounds can be quite wide. 

Instead, we assume that the potential outcomes under treatment for $A$ follow a specific parametric distribution while the potential outcomes for $B$ under treatment follow a different distribution. In our applications, we assume that the positive outcomes in $A$ and $B$ are Normally distributed with parameters $(\mu_{A1}, \sigma_{A1})$ and $(\mu_{B1}, \sigma_{B1})$. This results in a mixture model of two Normals for the positive outcomes\footnote{Technically, we should say non-zero outcomes, as the Normal distribution allows for negative outcomes. In our data, we do not observe any negative purchase amounts and the fitted Normal distributions have minimal negative support. We use the term ``positive outcomes'' to be consistent with the context and data.} in the treatment group (upper left in Figure \ref{fig:obs_groups}) with the likelihood: 
\begin{equation}
\ell(Y_i|Z_i=1, Y_i>0) = \pi_A \frac{1}{\sigma_{A1}}\phi\left(\frac{Y_i-\mu_{A1}}{\sigma_{A1}}\right) + \pi_B \frac{1}{\sigma_{B1}} \phi\left(\frac{Y_i-\mu_{B1}}{\sigma_{B1}}\right)
\label{eq:ll_treat_purch}
\end{equation}
where $\phi(\cdot)$ is the pdf of the standard Normal distribution.
By fitting a parametric mixture model to the data, the average outcomes and treatment effects for $A$ and $B$ can be identified. The distribution for the outcome need not be Normal as specified here; any distribution can be used, so long as the distributions are distinct so that the mixture model is identified. 

We specify likelihoods for the other three observational groups as well, which allows us to estimate $\pi_A$, $\pi_B$, $\pi_C$, $\tau_A$ and $\tau_B$ simultaneously. For the positive outcomes in the control group we specify another Normal model: 
\begin{equation}
\ell(Y_i | Z_i=0, Y_i>0) = \pi_A\frac{1}{\sigma_{A0}}\phi\left(\frac{Y_i-\mu_{A0}}{\sigma_{A0}}\right)
\label{eq:ll_ctrl_purch}
\end{equation}
Unlike Equation \eqref{eq:ll_treat_purch} where the parametric assumption is necessary for identification, the Normality assumption in Equation \eqref{eq:ll_ctrl_purch} is fairly innocuous, as the estimate of the mean is robust to misspecification. 

The zero outcomes (upper right and lower right in Figure \ref{fig:obs_groups}) occur with probabilities: 
\begin{equation}
\begin{aligned}
\ell(Y_i |Z_i=1, Y_i=0) = \pi_C\\
\ell(Y_i |Z_i=0, Y_i=0) = \pi_B + \pi_C
\end{aligned}
\label{eq:ll_zeros}
\end{equation}

Equations \eqref{eq:ll_treat_purch}, \eqref{eq:ll_ctrl_purch} and \eqref{eq:ll_zeros} specify a model for all of the experimental outcomes which can be used to obtain estimates of $\pi_A$, $\pi_B$, $\pi_C$, $\tau_A$ and $\tau_B$ using maximum likelihood estimation (MLE), method of moments or Bayesian inference. In our implementation, we use MLE with multiple starts, since the likelihood is not necessarily concave. No data is required beyond the treatment assignments $Z_i$ and the outcomes $Y_i$. Equation \eqref{eq:ll_zeros} identifies $\pi_A$, $\pi_B$ and $\pi_C$ and Equation \eqref{eq:ll_ctrl_purch} identifies $\mu_{A0}$ and $\sigma_{A0}$. Equation \eqref{eq:ll_ctrl_purch} also strengthens the identification of $\pi_A$ using the number of buyers in the control group. The mixture model in \eqref{eq:ll_treat_purch} is identified except in the pathological cases that $\mu_{A1} = \mu_{B1}$ and $\sigma_{A0}=\sigma_{B0}$,  or either $\pi_A$ or $\pi_B$ is zero; that is, the $A$ and $B$ strata must not be empty and there must be a difference in outcome distributions for treated customers in $A$ and $B$. 

As a function of the model parameters, the ATEs for the strata are: 
\begin{equation}
\tau_A = \mu_{A1}-\mu_{A0} \hspace{0.5in}
\tau_B = \mu_{B1} \hspace{0.5in}
\tau_C = 0
\label{eq:strata_ATEs}
\end{equation}
and the LS ATE is: 
\begin{align}
\tau^{\text{LS}} &= \pi_A (\mu_{A1} - \mu_{A0}) + \pi_B \mu_{B1}
\label{eq:LS_ATE}
\end{align}
Under MLE, a point estimate of the ATE can be computed from the point estimates for the model parameters following Equation \eqref{eq:LS_ATE}. Standard errors can be estimated by the delta method or by bootstrapping. Using the delta method, the estimated variance of the ATE is:
\begin{align}
    \widehat{Var}(\widehat{\tau}^{\text{LS}}) & = \left(\frac{\partial \widehat{\tau}^{\text{LS}}}{\partial \theta}\right)^T \times \widehat{V} \times \frac{\partial \widehat{\tau}^{\text{LS}}}{\partial \theta} \nonumber \\
    \frac{\partial\widehat{\tau}^{\text{LS}}}{\partial \theta} & = (\widehat{\mu}_{A1}-\widehat{\mu}_{A0},\widehat{\mu}_{B1}, \widehat{\pi}_A, -\widehat{\pi}_A, \widehat{\pi}_B, 0)^T
\end{align}
where $\theta = (\pi_A, \pi_B, \mu_{A1},\mu_{A0},\mu_{B1},\sigma)^T$ is the parameter vector, $\widehat{V}$ is the variance-covariance matrix of the parameters and $\frac{\partial\widehat{\tau}^{\text{LS}}}{\partial \theta}$ is the gradient of the estimated ATE. The delta method has the advantage that we can analytically derive the expressions for the gradient and the Hessian of the log-likelihood, which allow us to quickly and precisely compute the variance-covariance matrix. The R code for computing the gradient and the Hessian matrix is provided in Online Appendix \ref{app:r-code}. We verified the accuracy of the delta method approximation using simulation; across the simulations described in Section \ref{sec:precision} the average coverage is near-nominal and conservative at 0.958. 
The standard errors for the ATE can also be computed by bootstrapping, although this requires more computation.  

As we show in Section \ref{sec:precision}, the variance of the latent stratified estimator $\widehat{\tau}^{\text{LS}}$ would always be smaller than the variance of the standard DiM estimator in Equation \eqref{eq:diff-in-means_est}, in the hypothetical case where the strata membership for each customer was observed. In real data, this benefit is eroded somewhat because we have to estimate the LS model in Equations \eqref{eq:ll_treat_purch}-\eqref{eq:ll_zeros}. Despite the reduced benefit, we illustrate with a simulation study in Section \ref{sec:precision} that the sampling variation of $\widehat{\tau}^{\text{LS}}$ is substantially smaller than that of the DiM estimator $\widehat{\tau}^{\text{DiM}}$ over a wide range of true parameter values. 

\subsection{Relationship to principal stratification} 
Dividing customers into strata based on potential outcomes, as we do here, is called \emph{principal stratification} \citep{FrangakisRubin2002}. Latent stratification is a novel application of principal stratification where the strata are defined by potential outcomes for the focal response variable $Y_i$ (e.g. sales), and not potential outcomes for treatment compliance or study drop-out, which we do not address.\footnote{To avoid confusion, we do not use the strata labels ``Always Taker", ``Complier" and ``Never Taker", which evoke non-compliance.}  Appendix \ref{sec:principal_stratification} describes in detail how LS differs from other applications of principal stratification. 

While it may seem that latent stratification divides customers into strata based on the post-treatment response $Y_i$, it does not. In latent stratification the observed outcome ($Y_i$) \emph{partially reveals} the strata membership, which affects how we compute the likelihood. However, by definition, the treatment can not affect the potential outcomes for an experiment and the stratification is based on these potential outcomes. As \citet{FrangakisRubin2002} put it: 

\begin{quote} 
The key property of principal strata is that they are not affected by treatment assignment and therefore can be used just as any pretreatment covariate, such as age category. As a result, the central property of our principal effects is that they are always causal effects and do not suffer from the complications of standard posttreatment-adjusted estimands.
\end{quote}

\subsection{Assumptions of latent stratification and a misspecification test}
\label{sec:assumptions}
In addition to the standard assumptions for causal inference (unconfoundeness, SUTVA), LS makes two additional assumptions: (1) responses in $A$ and $B$ under treatment follow specific distributional forms and (2) there are no customers who purchase only if they are \emph{not} exposed to advertising. In our applications, we also assume that (3) $\sigma_{A0} = \sigma_{A1} = \sigma_{B1}$, which increases the identification of the mixture model, making the method more practical. 

Assumption (1) is unavoidable without pre-randomization covariates.\footnote{\cite{ding2011} show that a discrete covariate with $3$ levels identifies the average treatment effect for the $A$ group in a non-parametric model. Under specific assumptions, the proof can be extended to show that the ATE is non-parametrically identified with more levels.} Assumption (2) improves model identification. General two-component mixture models suffer from identification issues \citep{ho2022weak}. The latent stratification model is better-identified than a general mixture model because the size of the mixture components in Equation \eqref{eq:ll_treat_purch} are well-identified by the size of the observational groups (as discussed above). If we were to allow for a fourth stratum that purchases under control, but not treatment, all four observational groups (see Figure \ref{fig:obs_groups}) would comprise customers from two strata, requiring two mixture models, and the size of the strata would be less well-identified. We illustrate this by comparing the standard errors for the LS model to a more general two-component mixture model in Online Appendix \ref{sec:mixture_identification}. This monotonicity constraint is consistent with many other models of ad response which disallow negative response to marketing, e.g. ad stock models. One might think of this as using domain knowledge to improve identification \citep{kosyakova2023constrained}. However, the assumption might be violated if, for example, a promotion curtails consumer search by giving the impression that the assortment is unattractive or if advertising spillovers \citep{sahni2016advertising} lead to a reduction in sales for the advertised brand. As we show in Section \ref{sec:benchmark}, the assumption that the ad increases purchase incidence is not the primary driver of improvements in the precision of the ATE.  

Of course, there is some potential that the parametric assumptions we make are not justified by the data. To guard against this, we introduce a test of model misspecification that can be performed after estimating the latent stratification model. The test is based on the in-and-out-of-sample (IOS) test of \cite{presnell2004} that is motivated by cross-validation but provides formal inference. 

To perform the test, one removes each observation one-at-a-time, computes the MLE parameter estimates for the remaining data and computes the log-likelihood of the removed observation at those estimates. The sum of the log-likelihoods is then compared to the maximum log-likelihood with the complete data, and if they differ by much, this constitutes evidence of model misspecification, as removing one observation should not affect the log-likelihood by much.

Formally, if we denote the MLE parameter estimates as $\widehat{\theta}$, where in the LS model 
\newline $\theta = (\pi_A, \pi_B, \mu_{A1},\mu_{A0},\mu_{B1},\sigma)^T$, then the log-likelihood of the complete data can be written as $\sum_{i=1}^n \log \ell(Y_i| Z_i, \widehat{\theta})$. 

For each observation $i$, we denote by $\widehat{\theta}_{-i}$ the MLE parameter estimates when the data does not include $Y_i$, and write the likelihood of the $i$\textsuperscript{th} observation as: 
$\ell(Y_i| Z_i, \widehat{\theta}_{-i})$. The IOS statistic is defined as:
\begin{equation}
    \text{IOS} = \sum_{i=1}^n \log \ell(Y_i| Z_i, \widehat{\theta}) - \sum_{i=1}^n \log \ell(Y_i| Z_i, \widehat{\theta}_{-i})
\end{equation}

\cite{presnell2004} show that under a correctly specified model, when the number of observations $n \to \infty$, then $\text{IOS} \to |\theta|$, where $|\theta|$ is the number of model parameters (6 for LS).

Computation of the IOS statistic is similar to computing a jackknife estimator, which with a large sample can be very time consuming (as the likelihood needs to be maximized for every observation). \cite{presnell2004} recommend using an alternative parametric bootstrap method that hinges on the fact that in a correctly specified model the Fisher information matrix can be written either as $\mathbb{E}\left[\left(\frac{\partial}{\partial  \theta}\log \ell(Y|Z,\theta)\right)^2 | Z,\theta \right]$ or as 
$-\mathbb{E}\left[\frac{\partial^2}{\partial  \theta^2}\log \ell(Y|Z,\theta) | Z, \theta \right]$. 

To perform the test, we simulate 500 parametric bootstrap datasets from the latent stratification likelihood model using the MLE estimates obtained from the complete data. For each dataset $(Y^j, Z)$ of the 500 datasets we maximize the latent stratification likelihood and use the resulting MLE estimates $\widehat{\theta}^j$ to compute the sample analogs of the two expressions for the Fisher information matrices, which we denote as $A_n$ and $B_n$ and are defined as follows:
\begin{align}
    A_n^j & = -\frac{1}{n} \sum_i \frac{\partial^2}{\partial  \theta^2}\log \ell(Y_i^j| Z_i,\widehat{\theta}^j)\nonumber\\
    B_n^j & = \frac{1}{n} \sum_i \left(\frac{\partial}{\partial  \theta}\log \ell(Y_i^j|Z_i,\widehat{\theta}^j)\right)^2
\end{align}
The procedure results in 500 estimates of the IOS statistic which are computed as $\widehat{\text{IOS}}^j = tr\left((A_n^j)^{-1} B_n^j\right)$. The distribution of these 500 parametric bootstrap replicates constitutes an empirical distribution of the IOS statistic under the null hypothesis that the latent stratification model is correctly specified.
The distribution of these 500 bootstrap IOS values is compared to the IOS statistic computed over the complete dataset, and if the p-value is high, then we are unable to reject the null hypothesis of a correctly specified model.

\section{Example application}
\label{sec:applications}
\subsection{Data}

We estimate the ATE for 5 catalog incrementality tests that were conducted by a US multi-channel speciality retailer between September 2017 and February 2018. To illustrate how latent stratification works in varying situations, we estimate the latent stratification model separately for each of the five experiments. For each experiment, the retailer randomly selected approximately 140,000 customers from their active customer list and mailed a high-end catalog to half this list at random. The same customer may appear in multiple experiments, but this is immaterial as we analyze each experiment independently. For each customer, we observe all-channel purchases (net of returns) in the 30 days after the experiment. This was tracked using the retailer's regular name/address/email matching process.\footnote{Credit-card usage at this retailer is high and more than 80\% of transactions are matched to an existing customer in the CRM system.}

\begin{figure}[htbp]
\centering
\includegraphics[width=0.45\textwidth]{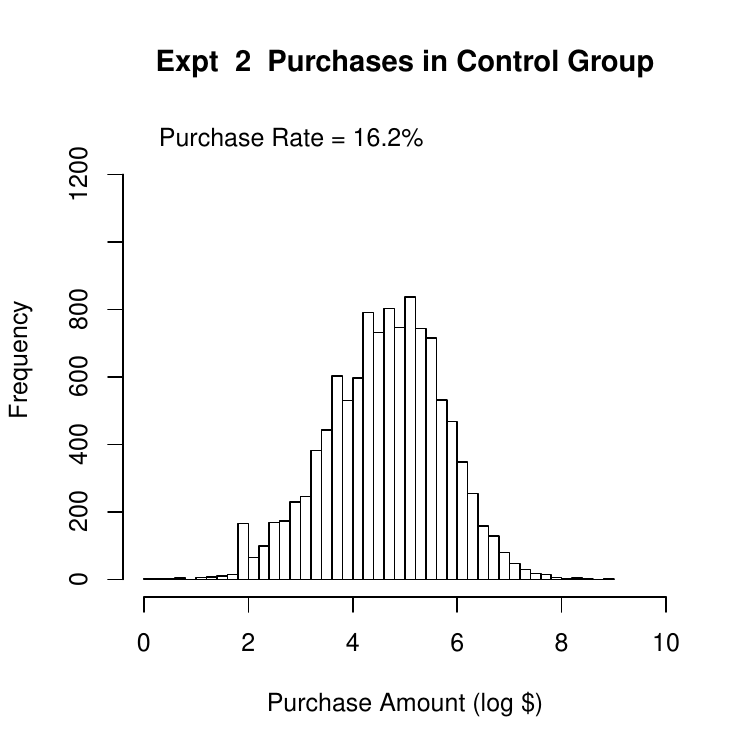}
\includegraphics[width=0.45\textwidth]{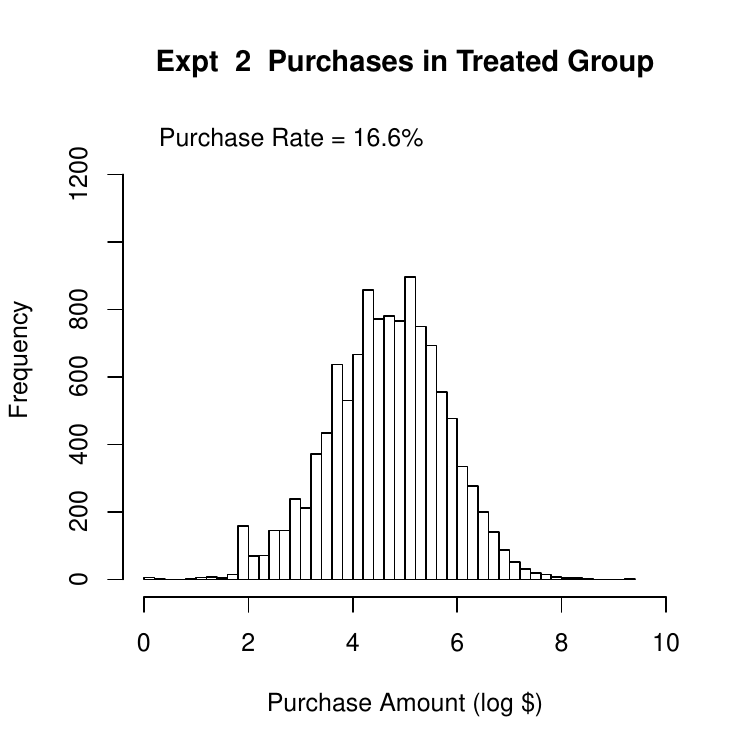}
\caption{Distribution of log-sales for Experiment 2}
\label{fig:expt2_hist}
\end{figure}

Table \ref{tab:expt_desc} contains basic summary statistics for each experiment. For the application the outcome $Y_i$ is log-sales, which we obtained by transforming the observed purchase amounts for each customer by adding 1 and taking the logarithm. Figure \ref{fig:expt2_hist} shows a histogram of the log-sales of positive purchases for Experiment 2. The histogram is consistent with the assumption that after the transformation, positive outcomes are Normally distributed in the control group and follow a mixture of two Normals in the treatment group. Histograms for other experiments are in Appendix \ref{sec:app1_hists}.
\linespread{1.1}
\begin{table}[htbp]
\caption{Summary of experiments} 
\label{tab:expt_desc}
\centering
\begin{tabular}{c l *{2}{S[table-format=2.3,
         input-decimal-markers={,},
         output-decimal-marker={,}]} *{4}{S[table-format=1.4,
         input-decimal-markers={.},
         output-decimal-marker={.}]}}
         \toprule
\multicolumn{1}{c}{Expt} & \multicolumn{1}{c}{Month} & \multicolumn{1}{c}{$n_1$} & \multicolumn{1}{c}{$n_0$} & \multicolumn{1}{>{\centering}m{0.8in}}{Avg. log-Sales Treated} & \multicolumn{1}{>{\centering}m{0.8in}}{Avg. log-Sales Control} & \multicolumn{1}{>{\centering}m{0.8in}}{Avg. Incidence Treated} & \multicolumn{1}{>{\centering}m{0.8in}}{Avg. Incidence Control}  \\
\midrule
1 & Sept 2017 & 69,291 & 68,990 & 4.605 & 4.575 & 0.164 & 0.165 \\ 
2 & Oct 2017 & 69,268 & 68,959 & 4.652 & 4.616 & 0.166 & 0.162\\ 
3 & Nov 2017 & 69,241 & 68,914 & 4.682 & 4.660 & 0.217 & 0.213\\ 
4 & Dec 2017 & 69,238 & 68,900 & 4.566 & 4.551 & 0.241 & 0.237\\ 
5 & Feb 2018 & 69,207 & 68,832 & 4.687 & 4.648 & 0.118 & 0.117\\ 
\bottomrule
\end{tabular}
\end{table}
\linespread{1.5}

\subsection{Estimates}
Estimated parameters for the latent stratification model for all 5 experiments are shown in Table \ref{tab:ps_wo_covs}. Taking Experiment 2 as an example, we estimate $\pi_A = 16.2\%$ and $\pi_B = 0.4\%$,  consistent with the observed purchase rates in Table  \ref{tab:expt_desc}. The average log-sales amount for treated customers in stratum $B$ is $\mu_{B1} = 2.992$, which corresponds to an average of $\exp(2.992 + 1.101^2/2)-1 = \$35.53$.\footnote{The mean of a log-Normal distribution with parameters $(\mu, \sigma)$ is $e^{\mu+\sigma^2/2}$.} For customers in $A$, we estimate the average treatment effect to be $\mu_{A1} - \mu_{A0} = 4.688-4.616 = 0.072$, which corresponds to an average lift of $\exp(4.688 + 1.101^2/2$) - $\exp(4.616+1.101^2/2$) = \$13.83. The model also allows us to disentangle the impact on the extensive and intensive margins of the catalog response: the estimated effect on the intensive margin is $\pi_A (\mu_{A1}- \mu_{A0}) = 0.162 \times (4.688-4.616) = 0.012$ and on the extensive margin it is $\pi_B \times \mu_{B1} = 0.004 \times 2.992 = 0.011$. Parameter estimates for other experiments are similar. 
\linespread{1.1}
\begin{table}[!htbp]
\caption{LS model parameter estimates for catalog experiments}
\label{tab:ps_wo_covs}
\centering
\begin{tabular}{l *{10}{c}}
\toprule
& \multicolumn{2}{c}{Expt 1} & \multicolumn{2}{c}{Expt 2} & \multicolumn{2}{c}{Expt 3} & \multicolumn{2}{c}{Expt 4} & \multicolumn{2}{c}{Expt 5}\\
 & est & se & est & se & est & se & est & se & est & se\\
\midrule
$\pi_A$ & 0.163 & 0.001 & 0.162 & 0.001 & 0.213 & 0.001 & 0.237 & 0.001 & 0.116 & 0.001 \\ 
$\pi_B$ & 0.003 & 0.001 & 0.004 & 0.001 & 0.004 & 0.001 & 0.005 & 0.002 & 0.003 & 0.001 \\ 
$\mu_{A0}$ & 4.575 & 0.010 & 4.616 & 0.010 & 4.660 & 0.009 & 4.551 & 0.008 & 4.648 & 0.012 \\ 
$\mu_{A1}$ & 4.635 & 0.015 & 4.688 & 0.015 & 4.715 & 0.013 & 4.597 & 0.012 & 4.731 & 0.018 \\ 
$\mu_{B1} $ & 2.946 & 0.224 & 2.992 & 0.232 & 2.995 & 0.215 & 3.214 & 0.201 & 3.022 & 0.216 \\ 
$\sigma$ & 1.091 & 0.006 & 1.101 & 0.006 & 1.088 & 0.005 & 1.061 & 0.005 & 1.100 & 0.007 \\ 
\bottomrule
\end{tabular}
\end{table}

The estimated within-stratum response variance for Experiment 2 is $\sigma^2 = (1.101)^2 = 1.212$, which is substantially smaller than the variance for the treatment and control groups overall, which are 3.198 and 3.094. This implies that the variances in the individual treatment effects within each stratum are substantially smaller than the variance across all customers, which is precisely the condition under which post-stratification produces a more precise estimate of the average treatment effect than the DiM estimator. 

The reduction in the variance of the ATE is substantial and can be seen in Table \ref{tab:app1_var_red}. For Experiment 2, the point estimates of the ATE are very similar ($\widehat{\tau}^{\text{LS}} = 0.0225$ versus $\widehat{\tau}^{\text{DiM}}=0.0247$), but the variance is reduced by 48.2\% from $(0.0095)^2=0.000090$ to $(0.0069)^2=0.000048$. Latent stratification increases the precision of the ATE estimate, giving decision makers a better estimate of the lift produced by the marketing treatment. 

\begin{table}[!htb]
\caption{Comparison of ATE estimates for catalog experiments}
\label{tab:app1_var_red}
\centering
\setlength\tabcolsep{3pt}
\begin{tabular}{p{1.4in}*{10}{c}}
\toprule
& \multicolumn{2}{c}{Expt 1} & \multicolumn{2}{c}{Expt 2} & \multicolumn{2}{c}{Expt 3} & \multicolumn{2}{c}{Expt 4} & \multicolumn{2}{c}{Expt 5}\\
 & est & se & est & se & est & se & est & se & est & se\\
\midrule
$\widehat{\tau}^{\text{DiM}}$ & 0.0040 & 0.0095 & 0.0247 & 0.0095 & 0.0245 & 0.0107 & 0.0248 & 0.0108 & 0.0110 & 0.0084\\
$\widehat{\tau}^{\text{LS}}$  & 0.0185 & 0.0061 & 0.0225 & 0.0069 & 0.0240 & 0.0074 & 0.0278 & 0.0091 & 0.0188 & 0.0055 \\ 
\midrule
\% Reduction Std Err & & 35.4 & & 28.0 & & 30.4 & & 16.3 & & 34.3 \\ 
\% Reduction Var & & 58.3 & & 48.2 & & 51.6 & & 30.0 & & 56.8 \\ 
\bottomrule
\end{tabular}
\end{table}

Table \ref{tab:app1_var_red} shows that for Experiments 2--4, the point estimates of the LS and DiM ATEs are generally similar. In Experiments 1 and 5 the LS estimate is higher. Latent stratification restricts the change in purchase incidence to be non-negative, which pushes the estimates up particularly for Experiments 1 and 5, where the observed purchase incidence in treatment and control are similar.\footnote{These two experiments were run during non-peak periods, while the other three were run during the holiday season, which may explain the difference in treatment effects.}  A shift in the estimated ATE may raise doubts about the validity of the latent stratification model and assumptions, which is why we test for misspecification in the next subsection.
 
\subsection{Misspecification test}

Table \ref{tb:ios} presents the results of the parametric bootstrap IOS test for the 5 experiments we analyze. The top row presents the IOS statistic for the complete experimental data, while the bottom row presents the p-values that compare the IOS statistic to the null distribution generated with 500 simulated parametric bootstraps. The top row shows that the IOS statistic for all experiments is quite close to 6, as expected, which provides evidence for the correctness of the latent stratification model. However, we observe that for Experiments 1 and 5, the p-values are quite small, suggesting rejection of the null hypothesis of correct model specification. Given such low p-values, we conclude that the latent stratification model is not correctly specified for Experiments 1 and 5. This conclusion is also consistent with the fact that purchase incidence for Experiment 1 is higher in the control group, which is in contradiction to the positive lift assumption of the latent stratification model. For Experiment 5, such misspecification might explain why the LS ATE estimates is higher than DiM.

\begin{table}[htbp]
    \centering
    \caption{IOS statistics for catalog experiments}
    \label{tb:ios}

    \begin{threeparttable}
    \begin{tabular}{c c c c c c}
    \toprule
    & Expt 1 & Expt 2 & Expt 3 & Expt 4 & Expt 5 \\
    \midrule
    $\widehat{IOS}(\widehat{\theta}^{\text{LS}})$ & 7.299 & 6.362 & 6.505 & 6.514 & 7.124 \\
    p-value & 0.038 & 0.238 & 0.138 & 0.106 & 0.022\\
    \bottomrule
    \end{tabular}
    \begin{tablenotes}[flushleft]
    \item The IOS statistics are computed using 500 parametric bootstrap draws. A small p-value (e.g. $<0.05$) indicates the model might be misspecified.
    \end{tablenotes}
   \end{threeparttable}
 
\end{table}

For the three other experiments (2, 3 and 4), we observe that all p-values are quite high. These higher p-values suggest that the latent stratification model is correctly specified for these experiments and one can use the LS ATE estimates for decision making.

\subsection{Value of a more precise ATE estimate}
\label{sec:decision}
Firms run incrementality experiments to determine whether to continue using a particular marketing channel. In this section, we illustrate how a more precise LS estimate leads to better decisions. Consider a firm who has estimated the ATE based on an incrementality experiment. If that firm is risk-neutral, then the optimal decision strategy is to continue with the marketing if the revenue and profit attributable to the marketing exceeds the cost. Otherwise, the firm should discontinue using that channel.\footnote{Firms also have the option to subtarget the audience, but here we focus on the binary decision to continue using the channel or not.}  The value of the LS ATE for decision making will depend on the context; if marketing effects are large relative to the noise, both estimators will perform well. But in typical marketing scenarios where the signal-to-noise ratio is low, the increased precision of LS is valuable for decision making.\footnote{We thank the Associate Editor for suggesting this decision frame.} 

To simulate the retailer repeatedly running catalog experiments, we draw a random sample of 25\% of observations from each experiment. We then estimate the ATE using both the DiM and LS estimators using the 25\% sample. If the estimated ATE is greater than zero, we assume the firm continues to use that marketing channel, otherwise it discontinues. For this illustration, we assume the cost of the marketing is zero.  The resulting profit of this decision is estimated from the remaining 75\% of the observations using an inverse-propensity weighted estimate \citep[similar to][]{hitsch2018heterogeneous, yoganarasimhan2022design, smith2022optimal}.\footnote{Since the experiment was randomized, this amounts to averaging the revenue for the remaining users in either the treatment or control group, and scaling up to the out-of-sample population size.} A decision error occurs when the firm discontinues marketing when the out-of-sample revenue is higher with marketing or vice versa. We repeat this analysis for 200 random samples, to estimate how this decision strategy performs in expectation.  

For Experiment 2, the DiM estimator results in an error rate of 7.5\%. This error is reduced to 0.5\% when using the LS estimator, resulting in an increase of 0.6\% in expected revenue. We find similar results for all five experiments (see Table \ref{tab:profit}).  

Somewhat surprisingly, the LS estimator performs especially well for Experiments 1 and 5, where the misspecification test suggests that the LS model is misspecified. For these two experiments, the LS estimate is potentially biased. This illustrates how a biased estimator with much lower sampling variance can result in improved decision making. That is, LS appears to make a favorable bias-variance trade-off. While the academic literature has held up experiments as the gold standard for obtaining unbiased estimates of ad effects, unbiasedness is not always necessary from a decision-making perspective. 

Another standard way to compare estimators is using the mean squared error (MSE) for out-of-sample prediction. The last row of Table \ref{tab:profit} reports the expected difference in out-of-sample MSE for DiM and LS. The difference in MSE is positive across all 5 experiments (although small) indicating that LS generally predicts out-of-sample outcomes better than DiM. 

\begin{table}[ht]
\centering
\caption{Performance of the DiM and LS estimators for prediction and decision making.}
\label{tab:profit}
\begin{threeparttable}
    \begin{tabular}{lrrrrr}
\toprule
& \multicolumn{1}{c}{Expt 1} & \multicolumn{1}{c}{Expt 2} & \multicolumn{1}{c}{Expt 3} & \multicolumn{1}{c}{Expt 4} & \multicolumn{1}{c}{Expt 5} \\
\midrule
DiM Error Rate & 30.0\% & 7.5\% & 11.5\% & 12.0\% & 28.5\% \\
LS Error Rate & 1.5\% & 0.5\% & 2.5\% & 2.5\% & 0.0\% \\
Increase in Expected Revenue & 2.0\% & 0.6\% & 0.4\% & 0.6\% & 2.3\% \\
\midrule
Exp. MSE Difference & 0.000004 &	0.000055 &	0.000068	& 0.000039	& 0.000042\\
\bottomrule
\end{tabular}
\begin{tablenotes}[flushleft]
    \item The MSE is estimated by predicting the average outcome for treatment and control in the held-out 75\% of the data for each random draw, and comparing it to the realized individual outcomes. The difference in MSEs is then averaged across the random draws. 
\end{tablenotes}
\end{threeparttable}
\end{table}

An alternative way to realize the benefits of reduced sampling variance is to reduce the sample size of a proposed test. The required sample size for a null hypothesis test scales with the variance, and for Experiment 2, this corresponds to a sample size for DiM that is nearly twice that required for LS. Details are provided in Online Appendix \ref{app:sample-sizes}. 

\subsection{Alternative ATE estimators}
\label{sec:benchmark}
This section compares the performance of LS to two alternatives for variance reduction -- zero-inflated models and models that use individual-level pre-randomization covariates.

A central feature of LS is that its structure accounts for the large number of zeros in the data. One might ask whether commonly used zero-inflated models might have similar benefits. We compare LS to a simpler model where the outcomes under treatment and control each follow different zero-inflated Normal distributions: 
\begin{align}\label{eq:zi-diff-in-norms}
\ell(Y_i|Z_i, Y_i>0) & = \left(\pi_1 \frac{1}{\sigma}\phi\left(\frac{Y_i-\mu_1}{\sigma}\right)\right)^{Z_i} \cdot 
\left(\pi_0 \frac{1}{\sigma}\phi\left(\frac{Y_i-\mu_0}{\sigma}\right)\right)^{1-Z_i}\nonumber\\ 
\ell(Y_i|Z_i, Y_i=0) & = (1-\pi_1)^{Z_i} \cdot (1-\pi_0)^{1-Z_i}
\end{align}
and construct an estimator of the ATE as $\widehat{\tau}^{\text{ZI}} = \widehat{\pi}_1 \cdot \widehat{\mu}_1 - \widehat{\pi}_0 \cdot \widehat{\mu}_0$. We estimate this model for the example experiments by MLE and estimate its sampling variance using bootstrapping. Since latent stratification does not allow for a decrease in purchase incidence with treatment, we also estimate a constrained version of this model where $\pi_1 \ge \pi_0$ and refer to the corresponding estimator as $\widehat{\tau}^{\text{ZI+}}$. These models account for the large number of zeros, but the estimator is constructed like a DiM estimator without stratification. Table \ref{tab:app_post_strat} shows that this does not provide any benefit in terms of the variance of the ATE estimate; both estimators have sampling variation that is similar to DiM. The only exception is that $\tau^{\text{ZI+}}$ has lower sampling variance than DiM in Experiments 1 and 5 where the data is inconsistent with the ZI+ assumption of a positive lift in incidence. Simply accounting for zeros in the model is insufficient to reduce the variance substantially.

\begin{table}[!htb]
\caption{Comparison of ATE estimates for catalog experiments}
\label{tab:app_post_strat}
\centering
\scriptsize
\setlength\tabcolsep{3pt}
\begin{tabular}{p{1.2in}*{10}{c}}
\toprule
ATE Estimate & \multicolumn{2}{c}{Expt 1} & \multicolumn{2}{c}{Expt 2} & \multicolumn{2}{c}{Expt 3} & \multicolumn{2}{c}{Expt 4} & \multicolumn{2}{c}{Expt 5}\\
 & est & se & est & se & est & se & est & se & est & se\\
\midrule
$\widehat{\tau}^{\text{DiM}}$ & 0.0040 & 0.0095 & 0.0247 & 0.0095 & 0.0245 & 0.0107 & 0.0248 & 0.0108 & 0.0110 & 0.0084\\
\midrule
\multicolumn{3}{l}{\textbf{Alternative Zero-inflated Models}}\\
$\tau^{\textnormal{ZI}}$ & 0.0040 & 0.0096 & 0.0247 & 0.0096 & 0.0245 & 0.0109 & 0.0248 & 0.0108 & 0.0110 & 0.0088\\
$\tau^{\textnormal{ZI+}}$& 0.0050 & 0.0054 & 0.0247 & 0.0096 & 0.0245 & 0.0103 & 0.0248 & 0.0107 &  0.0110 & 0.0069\\
\midrule
\multicolumn{3}{l}{\textbf{Covariate Adjustments}} \\
Regression Adjustment & 0.0044 & 0.0086 & 0.0252 & 0.0087 & 0.0256 & 0.0097 & 0.0257 & 0.0099 & 0.0112 & 0.0078   \\
Causal Forest (AIPW) & 0.0029 & 0.0085 & 0.0246 & 0.0086 & 0.0241 & 0.0097 & 0.0257 & 0.0099 & 0.0108 & 0.0078\\
Causal Forest (TMLE) & 0.0029 & 0.0086 & 0.0246 & 0.0087 & 0.0241 & 0.0097 & 0.0258 & 0.0099 & 0.0108 & 0.0078 \\
MLRATE (XGBoost) & 0.0037 & 0.0084 & 0.0235 & 0.0085 & 0.0247 & 0.0095 & 0.0254 & 0.0097 & 0.0111 & 0.0076 \\
\midrule
$\widehat{\tau}^{\text{LS}}$  & 0.0185 & 0.0061 & 0.0225 & 0.0069 & 0.0240 & 0.0074 & 0.0278 & 0.0091 & 0.0188 & 0.0055 \\ 
\bottomrule
\end{tabular}
\end{table}

A second popular approach for reducing the sampling variance of the ATE is to use pre-randomization covariates for post-stratification or regression adjustment \citep[e.g.][]{deng2013}.\footnote{Regression adjustment is equivalent to post-stratification when the covariate is discrete and is interacted with the treatment indicator.} In our example application, the retailer maintains a CRM system recording previous transactions for each customer. Using this data, we computed several RFM-style features summarizing customer's relationship to the retailer from the customer's first purchase to the start of Experiment 1: months since last transaction (R), average transactions per month (F), amount spent (net returns) per month (M), amount spent per month when catalog is not sent (C), number of times customer was targeted with a catalog over the prior 13 months (T), the difference in purchase amount between months when the customer received a catalog and did not (D) and the difference in frequency of purchases between months when the customer received a catalog and did not (I). These variables are all likely to be related to the individual-level treatment effects and thus are attractive pre-randomization covariates for regression adjustment. 

We used these covariates in a linear regression relating the purchase amount in the experiment to the treatment indicator, the covariates and the two-way interactions between those. If the covariates are correlated with the individual treatment effects, the regression will produce an estimate of the overall ATE with a smaller standard error. Across all five experiments, we find that it does reduce the standard error, but not nearly as much as latent stratification (see Table \ref{tab:app_post_strat}). For example, in Experiment 2, the regression adjustment reduced the standard error from 0.0095 to 0.0087 or 8.4\%, while LS reduces it to 0.0069 or 27.4\%. The full set of linear regression coefficients are shown in the Online Appendix \ref{app:regression_adjustment}. 

One limitation of regression adjustment is that it assumes a linear relationship between the covariates and the outcome.\footnote{Some analysts will dichotomize the covariates to avoid this assumption and maintain the unbiasedness of the ATE estimate.} An alternative is to use a more flexible machine learning model like a causal forest to estimate heterogeneous treatment effects as a function of the covariates. Then, the ATE is estimated using an augmented inverse-probability weighted estimator (AIPW) or targeted maximum likelihood estimation (TMLE) \citep{wager2018estimation, tibshirani2018package}. We implemented this and the resulting causal forest estimates of the ATEs have a similar standard errors as the regression adjustment estimate. Another approach is to use machine learning predictors of the outcome in regression adjustment to reduce estimator variance, which generalizes linear regression adjustment and allows for complex interactions between covariates. We implemented the MLRATE method of \cite{guo2021machine} using XGBoost \cite{chen2016} for predicting outcomes. The method performs slightly better than linear regression adjustment in reducing variance, but only marginally. Thus, even when using the most sophisticated methods, the benefit of pre-randomization covariates is dominated by LS for this application. Of course, with different covariates or in different applications, there is always potential for regression adjustment to produce greater benefits when there is a covariate available that is highly-correlated with the individual treatment effects.  We also note that using covariates has several down-sides: 1) the pre-randomization variables require tracking, which is less privacy-friendly, 2) it can be dependant on the model specification, e.g. linear regression, and 3) it will only reduce the sampling variance if the covariates are correlated with the outcome.

\section{When is latent stratification beneficial?}
\label{sec:precision}
The previous section shows that LS performs well in the example catalog experiments. In this section, we show that latent stratification provides a reduction in sampling variation across a broader range of scenarios. The improved precision of the LS estimator is driven primarily by two factors. First, as in the case of standard post-stratification, the estimate is more precise when the expected outcomes in each strata are different. 

Second, because strata membership are latent (unobserved), the mixture model in Equation \eqref{eq:ll_treat_purch} might not be well-identified which adds additional variability to our estimate of the overall ATE, $\widehat{\tau}^{\text{LS}}$. To understand the impact of these two factors and provide intuition for when LS is expected to perform best, we proceed with an analysis of the benefit of LS when strata membership are observed, which allows us to derive a closed-form equation for the benefit. We show that the variance of the LS estimator would \emph{always} be lower than the DiM estimator. Then, to account for the additional sampling variation introduced by the estimation of the mixture model, we present a simulation study for a range of data sets that are typical of incrementality experiments in marketing. Empirically, the method reduces variance the most when the mixture model is well-identified.

\subsection{Theoretical benefit in an oracle scenario}
\label{sec:known}
In the hypothetical case where strata membership are known, we can compute the variance of the estimator in \eqref{eq:LS_ATE} in closed form. We call this the \emph{oracle scenario} and use it as a benchmark to provide intuition about the source of the benefit from stratifying. 

In the oracle scenario, each customer in the experiment can be represented by the tuple $(Y_i, Z_i, X_{iA}, X_{iB}, X_{iC})$ where $Y_i$ and $Z_i$ are the observed outcome and treatment assignment as before, and $X_{iA}$, $X_{iB}$, $X_{iC}$ are indicators for the strata membership of customer $i$. For example, customer $i$ in stratum $A$ would have $X_{iA}=1$, $X_{iB}=0$ and $X_{iC}=0$. This notation allows us to derive closed-form expressions for the maximum likelihood estimators $\widehat{\pi}_A$, $\widehat{\pi}_B$, $\widehat{\mu}_{A1}$, $\widehat{\mu}_{A0}$ $\widehat{\mu}_{B1}$, and $\widehat{\sigma}$ (details appear in Appendix \ref{app:benefit}). The delta method allows us to compute the variance of the stratified estimator $\widehat{\tau}^{\text{LS}}$, which we then compare to the variance of the difference-in-means estimator $\widehat{\tau}^{\text{DiM}}$ in Equation \eqref{eq:diff-in-means_var} and find the following:
\begin{prop}
\label{prop:strat-benefit}
When $\sigma_{A1}=\sigma_{A0}=\sigma_{B1}=\sigma$, then the variance of the latent stratification estimator is always smaller than the variance of the difference-in-means estimator. Further, the difference equals:
\begin{align}
    Var(\widehat{\tau}^{\text{DiM}})-Var(\widehat{\tau}^{\text{LS}}) & =  \frac{\pi_A \mu_{A0}^2 -(\pi_A \mu_{A0})^2}{n_0}+\frac{\pi_A \mu_{A1}^2 + \pi_B\mu_{B1}^2 -(\pi_A \mu_{A1} + \pi_B \mu_{B1})^2}{n_1} \nonumber \\
        & -\left(\frac{\pi_A(\mu_{A1}-\mu_{A0})^2 +\pi_B\mu_{B1}^2 -\left(\pi_A (\mu_{A1}-\mu_{A0})+\pi_B \mu_{B1}\right)^2}{n}\right)\label{eq:ls-benefit}
\end{align}
\end{prop}

A somewhat surprising implication of \eqref{eq:ls-benefit} is that the benefit of latent stratification does not depend on the variance within each stratum, $\sigma^2$. The variance of the LS estimator can be separated into the variance between the strata and within the strata, and the variance within the strata is the same for $Var(\widehat{\tau}^{\text{DiM}})$ and $Var(\widehat{\tau}^{\text{LS}})$, and cancels out (see Appendix \ref{app:benefit} for details). 

We can also simplify the expression in \eqref{eq:ls-benefit} by assuming that $n_0=n_1=n/2$ and rewrite it as:
\begin{align}
    Var(\widehat{\tau}^{\text{DiM}})-Var(\widehat{\tau}^{\text{LS}}) & =  \frac{\pi_A \left(\frac{\mu_{A0}}{2}+\frac{\mu_{A1}}{2}\right)^2+\pi_B(\frac{\mu_{B1}}{2})^2-\left(\pi_A \left(\frac{\mu_{A0}}{2}+\frac{\mu_{A1}}{2}\right)+ \pi_B \frac{\mu_{B1}}{2}\right)^2}{\frac{n}{4}}\label{eq:ls-benefit-simple}
\end{align}
This expression has the form of a variance\footnote{We use the classic formulation $Var(X) = \mathbb{E}[X^2]-\mathbb{E}[X]^2$.} of a categorical random variable with three outcomes $(\frac{\mu_{A0}+\mu_{A1}}{2}, \frac{\mu_{B1}}{2}, 0)$ that occur with probabilities $(\pi_A, \pi_B, 1-\pi_A-\pi_B)$. These three outcomes are the expected values of $Y_i$ within each stratum, averaging over the treatment assignment probabilities. When this variance is maximized, the benefit from latent stratification is maximal. 

Note that the benefit does not come simply from separating out stratum $C$ from the other two strata. If that was the case, one could construct an estimator based on stratifying customers into two strata: those who will not buy regardless of treatment and all others. The analysis of this model (provided in Appendix \ref{app:two-strata}) shows that this estimator is equivalent to the difference-in-means estimator and therefore does not have lower variance than DiM. This model is closely related to the zero-inflated model estimated for the example application (Section \ref{sec:benchmark}), which also has similar variance to DiM. 

Because the strata are never actually observed, Proposition \ref{prop:strat-benefit} provides an upper-bound on the benefit that can be achieved by latent stratification. Comparing the variance of $\widehat{\tau}^{\text{LS}}$ in the oracle scenario to the realized variance in our empirical application provides a sense of how much precision is lost because the strata are latent. For Experiment 2 the empirical standard error is estimated at 0.0069, while plugging into the expression for $Var(\widehat{\tau}^{\text{LS}})$ (see Appendix \ref{app:benefit}) yields a standard error of 0.0025. Thus, the LS estimate may have substantially larger variance than the oracle scenario. In the next section, we empirically explore the variance reduction for latent stratification when the strata are unknown for a wider range of parameter values. 

\subsection{Empirical reduction in variance}\label{sec:unknown}

When strata are unknown, there are no closed form expressions for the MLE estimates of $\widehat{\mu}_{A1}$ and $\widehat{\mu}_{B1}$, hence we resort to simulation to understand how LS improves the sampling variance over DiM. To estimate the sampling variation, we generate 2000 data sets of size $n=100,000$ from the LS model using parameter values of $\pi_A=0.16$, $\pi_B=0.01$, $\pi_C=0.83$, $\mu_{A1}=4.7$, $\mu_{A0}=4.5$, $\mu_{B1}=3$ and $\sigma_{A1} = \sigma_{A0} = \sigma_{B1} = 1$. These values are similar to the point estimates for Experiment 2 in Section \ref{sec:applications}. We then compute $\widehat{\tau}^{\text{LS}}$ and $\widehat{\tau}^{\text{DiM}}$ and the oracle LS estimate for each of the 2000 simulated data sets and compare the estimates to the known truth.

Table \ref{tab:bias_var} shows that the estimated bias is numerically zero for $\widehat{\tau}^{\textnormal{DiM}}$ and the oracle estimator, as expected; both of these estimators are unbiased. $\widehat{\tau}^{\textnormal{LS}}$ is, like all MLE estimates, asymptotically unbiased when the model is correctly specified. Table 7 confirms that a sample size of 100,000 is sufficient to reach a nearly-unbiased estimate; the estimated bias for $\widehat{\tau}^{\textnormal{LS}}$ is within the noise of the simulation. More importantly, the simulation confirms that the sampling variation is substantially smaller for $\widehat{\tau}^{\textnormal{LS}}$, with a variance of 0.0000700 versus 0.0001257 (44\% reduction). The sampling variation dominates the bias resulting in an MSE that is much lower for $\widehat{\tau}^{\textnormal{LS}}$ versus $\widehat{\tau}^{\textnormal{DiM}}$. Even if the $\widehat{\tau}^{\textnormal{LS}}$ has a small finite-sample bias, the result is a favorable bias-variance trade-off. More details on the accuracy of $\widehat{\tau}^{\text{LS}}$ under both correct and incorrect specifications are provided in Online Appendix \ref{sec:accuracyLSATE}.

\begin{table}[h]
    
\caption{Simulated bias, variance and mean squared error of LS and DiM estimates for a synthetic data set.}
\label{tab:bias_var}
\centering
\begin{threeparttable}

\begin{tabular}{@{\extracolsep{5pt}} ccccc} 
\toprule
 & bias & $\textnormal{bias}^2$ & var & MSE \\ 
\midrule
$\widehat{\tau}^{\textnormal{DiM}}$ & $0.0000$ & $0.0000$ & $0.0001257$ & $0.0001256$ \\ 
$\widehat{\tau}^{\textnormal{LS}}$ & $0.0002$ & $0.00000004$ & $0.0000700$ & $0.0000700$ \\ 
Oracle & $0.0000$ & $0.0000$ & $0.0000065$ & $0.0000065$ \\ 
\bottomrule
\end{tabular} 
\begin{tablenotes}[flushleft]
\item Estimates based 2000 sets generated from the LS model with parameters  $\pi_A=0.16$, $\pi_B=0.01$, $\pi_C=0.83$, $\mu_{A1}=4.7$, $\mu_{A0}=4.5$, $\mu_{B1}=3$ and $\sigma_{A1} = \sigma_{A0} = \sigma_{B1} = 1$ where the true ATE is 0.4862.    
\end{tablenotes}
\end{threeparttable}

\end{table}

\begin{figure}[!htb]
\centering
\includegraphics[page=11, width=0.49\textwidth]{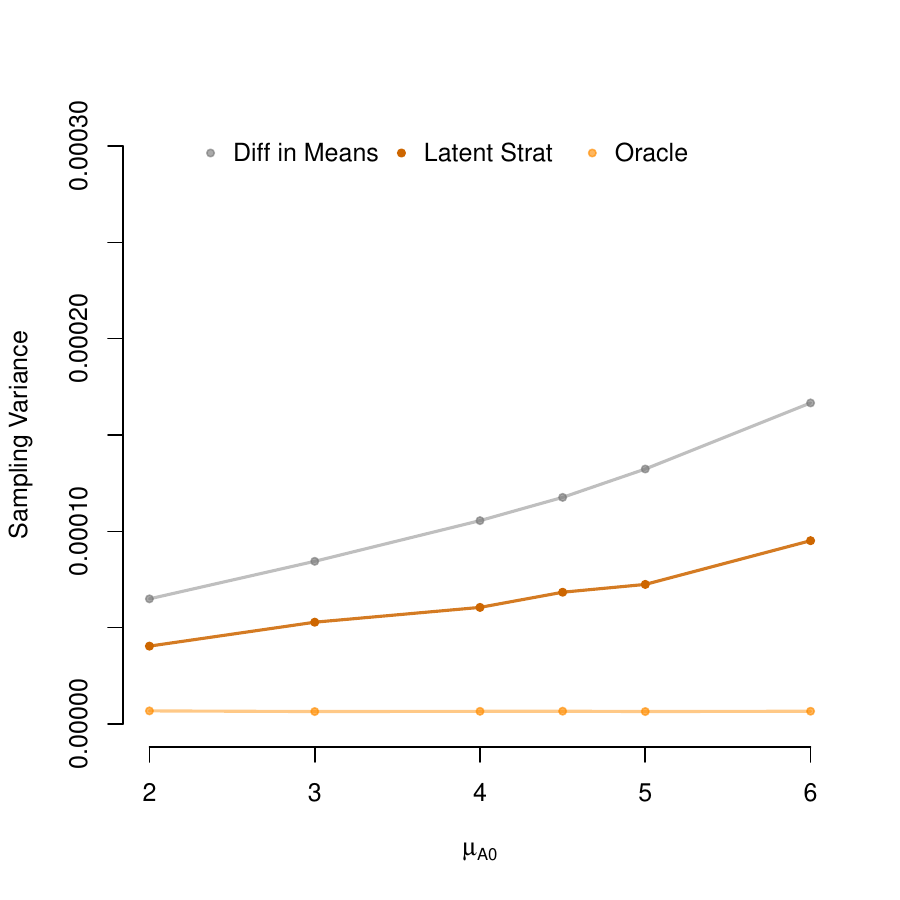}
\includegraphics[page=17, width=0.49\textwidth]{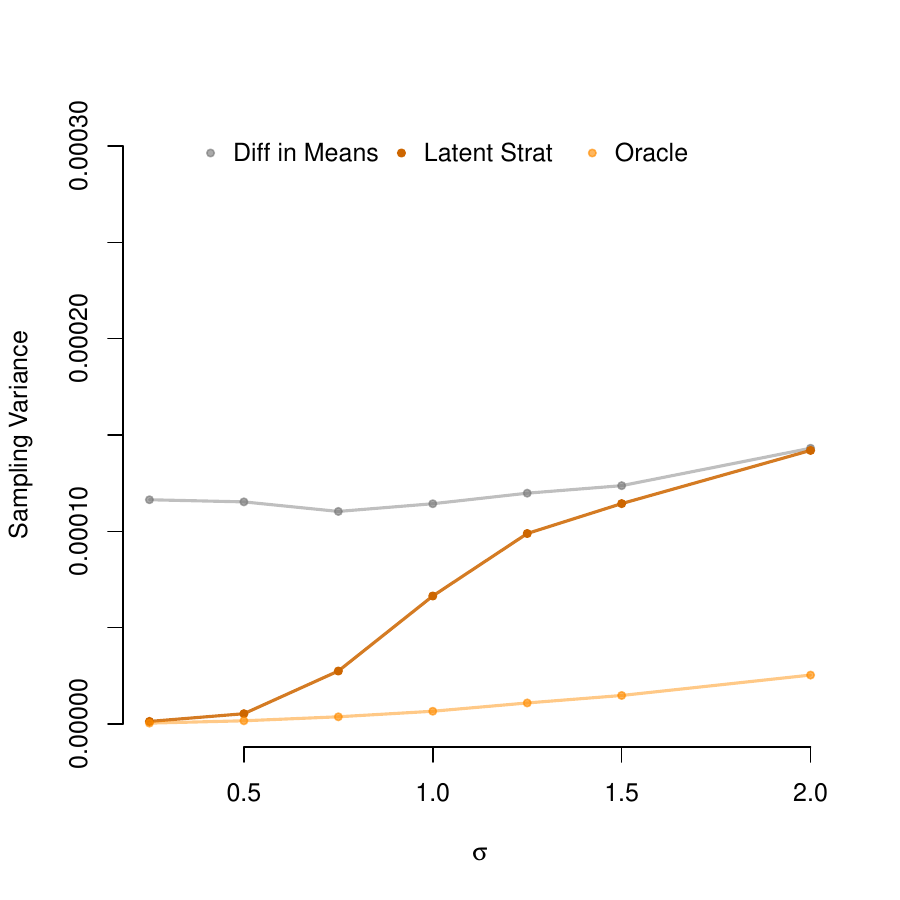}
\caption{Sampling variation in alternative estimators of the ATE for different values of $\mu_{A0}$ and $\sigma$. Other parameters are fixed at $\pi_A=0.16$, $\pi_B=0.01$, $\pi_C=0.83$, $\mu_{A1}=4.7$, $\mu_{A0}=4.5$, $\mu_{B1}=3$ and $\sigma_{A1} = \sigma_{A0} = \sigma_{B1} = 1$.}
\label{fig:sensitivity_sigmamuA0}
\end{figure}

\begin{figure}[!htb]
\centering
\includegraphics[page=8, width=0.49\textwidth]{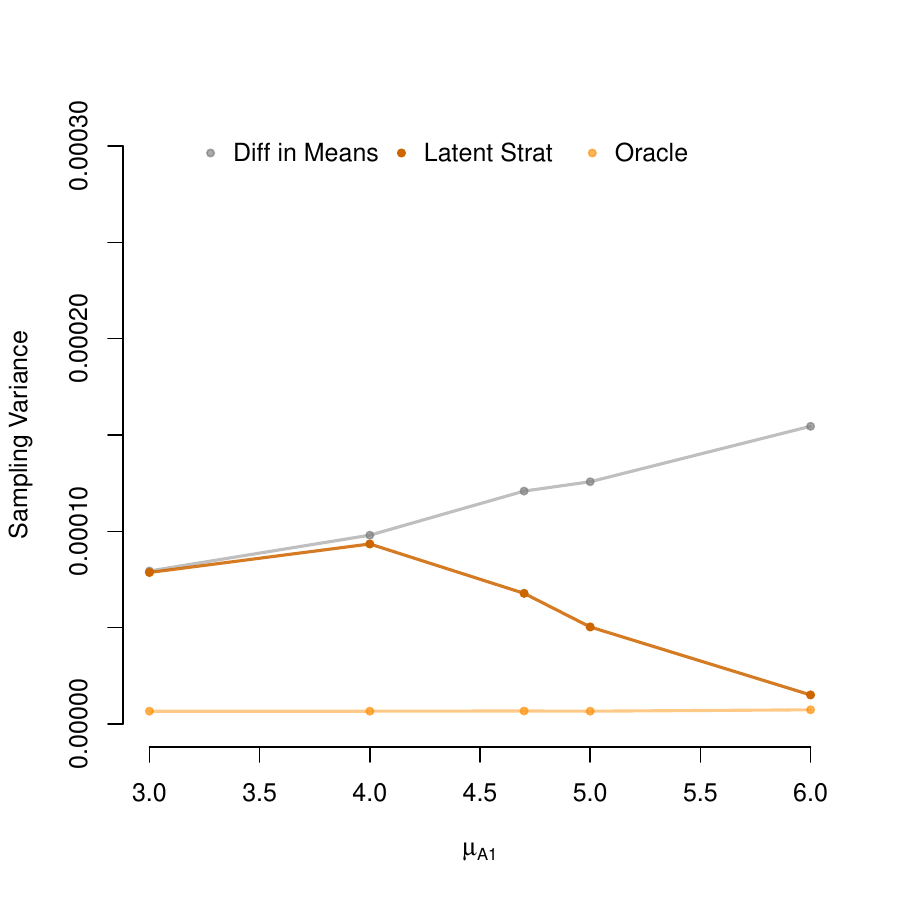}
\includegraphics[page=14, width=0.49\textwidth]{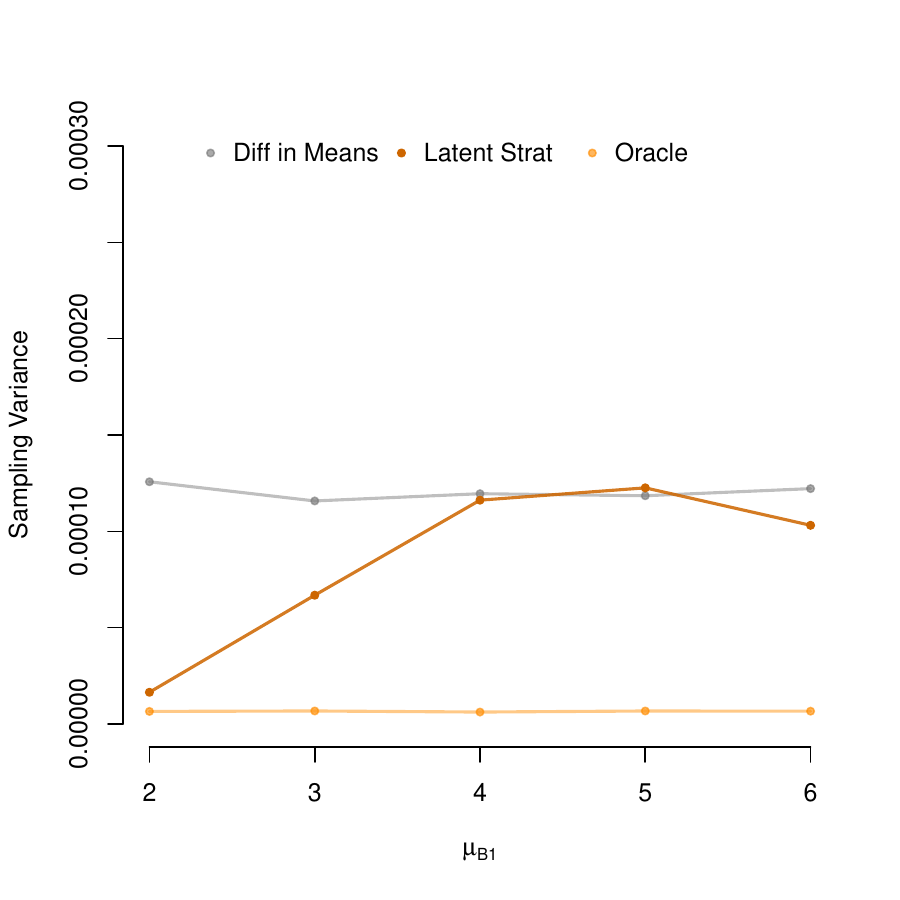}
\caption{Sampling variation in alternative estimators of the ATE for different values of $\mu_{A1}$ and $\mu_{B1}$. Other parameters are fixed at $\pi_A=0.16$, $\pi_B=0.01$, $\pi_C=0.83$, $\mu_{A1}=4.7$, $\mu_{A0}=4.5$, $\mu_{B1}=3$ and $\sigma_{A1} = \sigma_{A0} = \sigma_{B1} = 1$.} 
\label{fig:sensitivity_muA1muB1}
\end{figure}

\begin{figure}[!htb]
\centering
\includegraphics[width=0.49\textwidth, page=4]{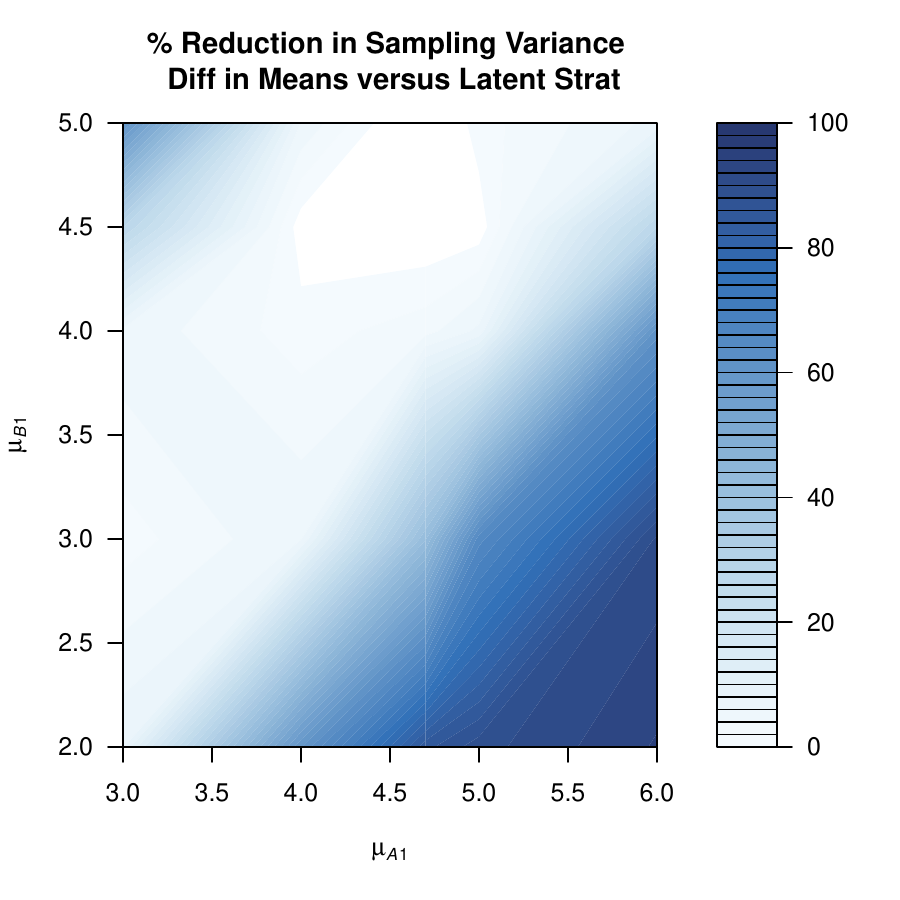}
\caption{Percent reduction in sampling variation $\left(1-\frac{Var(\widehat{\tau}^{\text{LS}})}{Var(\widehat{\tau}^{\text{DiM}})}\right)$ for varying values of $\mu_{A1}$ and $\mu_{B1}$. Other parameters are fixed at $\pi_A=0.16$, $\pi_B=0.01$, $\pi_C=0.83$, $\mu_{A0}=4.5$ and $\sigma_{A1} = \sigma_{A0} = \sigma_{B1} = 1$.}
\label{fig:sensitivity_mus}
\end{figure}

\begin{figure}[!htb]
\centering
\includegraphics[page=5, width=0.49\textwidth]{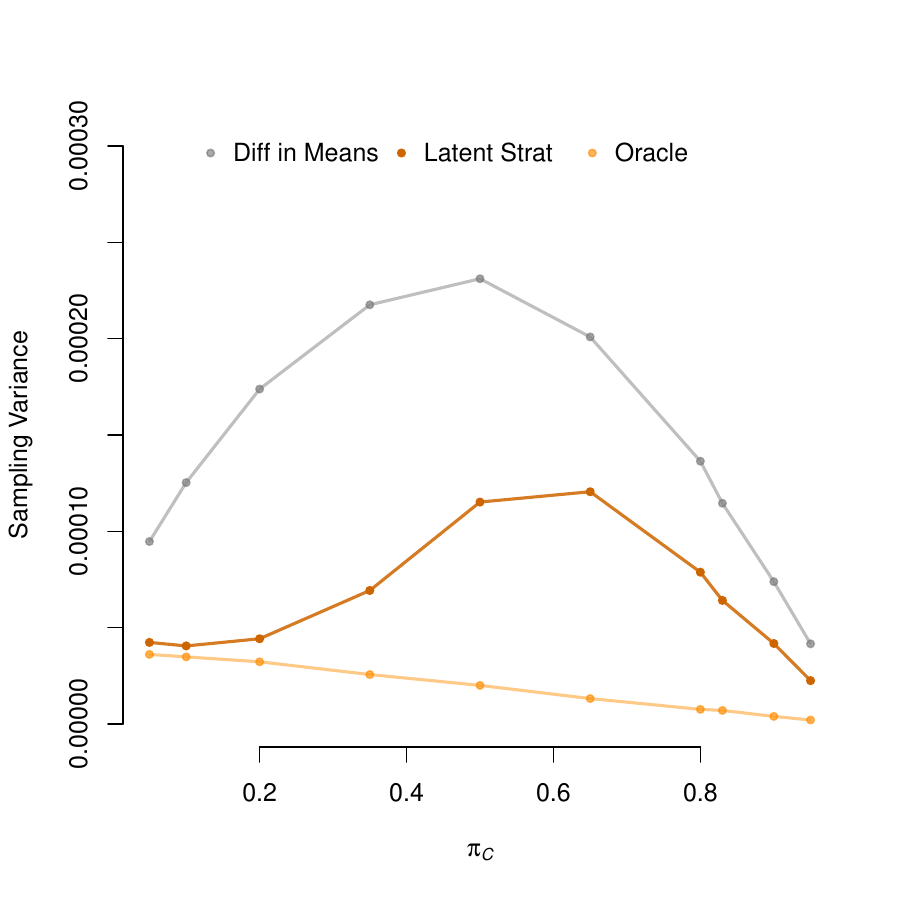}
\includegraphics[page=2, width=0.49\textwidth]{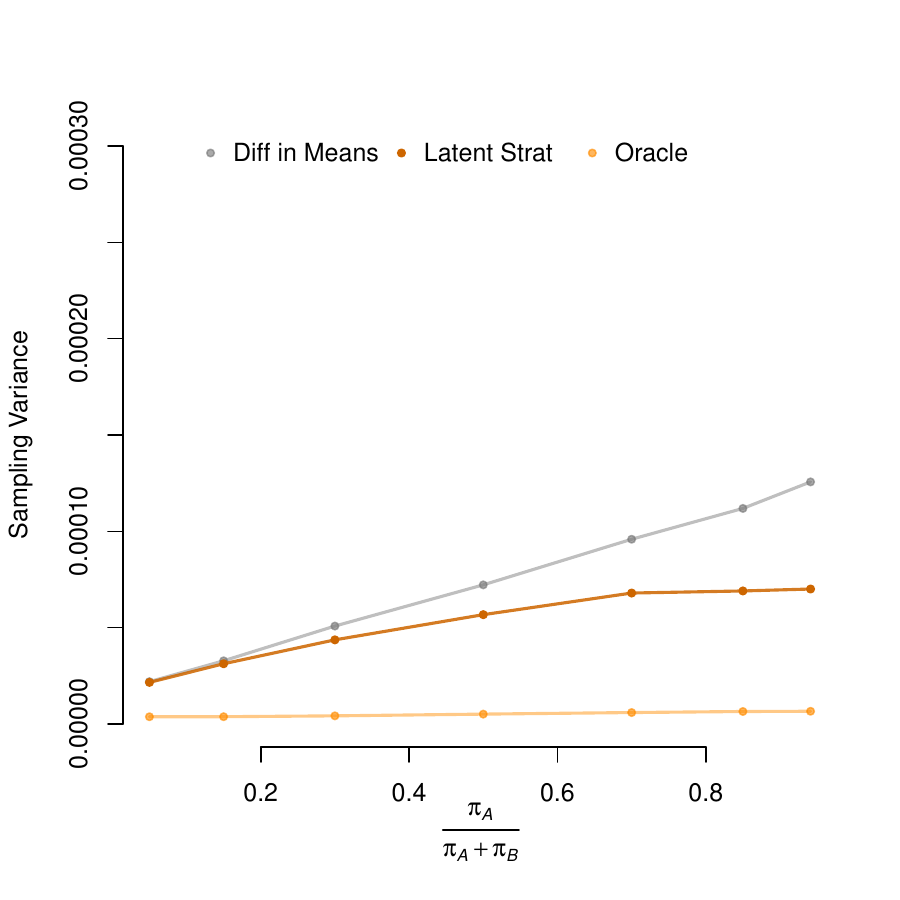}
\caption{Sampling variation in alternative estimators of the ATE for different values of $\pi_C$ and $\frac{\pi_A}{\pi_A + \pi_B}$. As $\pi_C$ is varied, $\frac{\pi_A}{\pi_A + \pi_B}$ is fixed at 0.16/(0.16+0.01). As $\frac{\pi_A}{\pi_A + \pi_B}$ is varied $\pi_C$ is fixed at 0.83. Other parameters are fixed at $\mu_{A1}=4.7$, $\mu_{A0}=4.5$, $\mu_{B1}=3$ and $\sigma = 1$.}
\label{fig:sensitivity_pis}
\end{figure}

\begin{figure}[!htb]
\centering
\includegraphics[page=4, width=0.49\textwidth]{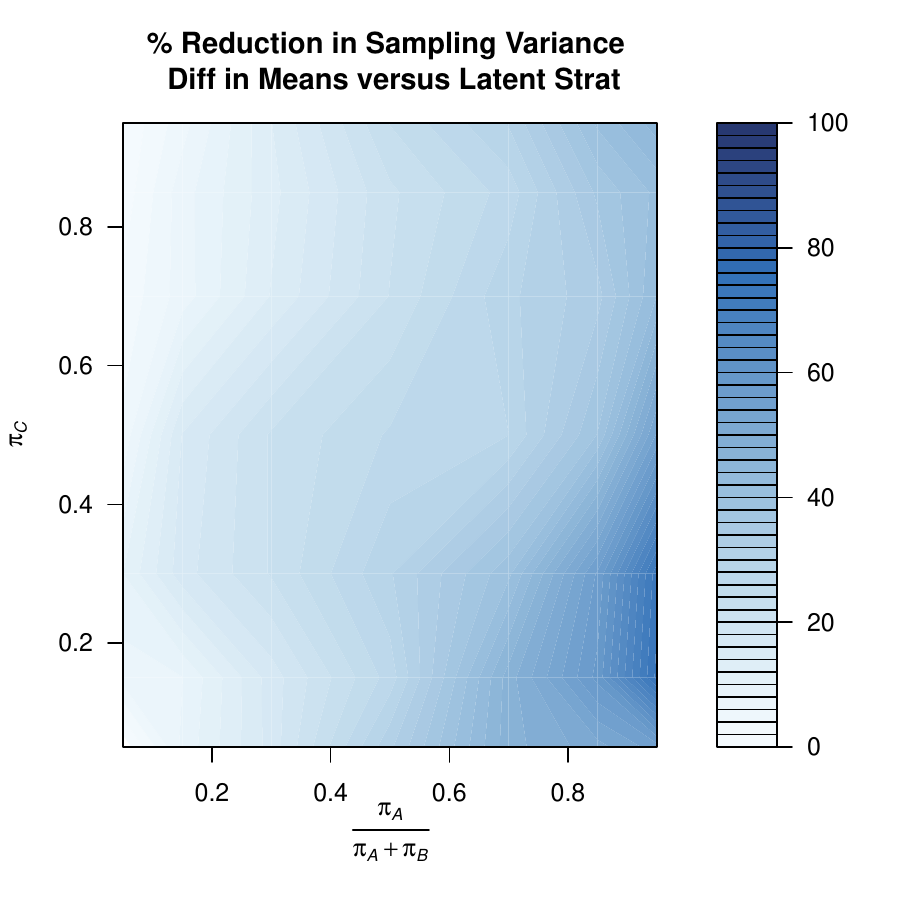}
\caption{Percent reduction in sampling variation $\left(1-\frac{Var(\widehat{\tau}^{\text{LS}})}{Var(\widehat{\tau}^{\text{DiM}})}\right)$ for different values of $\pi_C$ and $\frac{\pi_A}{\pi_A+\pi_C}$. As $\pi_C$ is varied, $\frac{\pi_A}{\pi_A + \pi_B}$ is fixed at 0.16/(0.16+0.01). As $\frac{\pi_A}{\pi_A + \pi_B}$ is varied $\pi_C$ is fixed at 0.83. Other parameters are fixed at $\mu_{A1}=4.7$, $\mu_{A0}=4.5$, $\mu_{B1}=3$ and $\sigma = 1$.}
\label{fig:sensitivity_pis2}
\end{figure}

To provide insight into when $\widehat{\tau}^{\text{LS}}$ has reduced sampling variance versus $\widehat{\tau}^{\text{DiM}}$, we vary the true value of each parameter one-at-a-time holding the other parameters fixed at the baseline values. We then generate 2000 data sets for each set of parameters and estimate the ATE using both estimators. The sampling variance of these estimators is plotted in Figures \ref{fig:sensitivity_sigmamuA0}--\ref{fig:sensitivity_pis2}. The charts also include the sampling variance under the oracle scenario, which provides a lower-bound on the sampling variation of the LS ATE. 

Figure \ref{fig:sensitivity_sigmamuA0} shows the sampling variation of the ATE estimators as $\mu_{A0}$ and $\sigma$ are varied. The left panel of Figure \ref{fig:sensitivity_sigmamuA0} shows that the LS estimator consistently has proportionally lower sampling variance than the DiM estimator across different values of $\mu_{A0}$. The right panel of Figure \ref{fig:sensitivity_sigmamuA0} shows that when $\sigma$ is low, $\widehat{\tau}^{\text{LS}}$ has nearly the same sampling variance as the oracle estimator. As $\sigma$ increases, $Var(\widehat{\tau}^{\text{LS}})$ increases, due to the increased difficulty of estimating the mixture model in Equation \ref{eq:ll_treat_purch}. At $\sigma=2.0$ the sampling variance of LS is nearly the same as DiM. 

Figure \ref{fig:sensitivity_muA1muB1} shows the sampling variation of the ATE estimators as $\mu_{A1}$ and $\mu_{B1}$ are varied. These two parameters represent the means of the two mixture components among users for which $Z_i=1$ and $Y_i>0$. When $\mu_{A1}$ and $\mu_{B1}$ have similar values, the mixture model becomes weakly-identified, leading to more sampling variation in $\widehat{\tau}^{\text{LS}}$. For example, in the left panel of Figure \ref{fig:sensitivity_muA1muB1}, when $\mu_{A1}$ is closer to 3 (the baseline value of $\mu_{B1}$), the LS sampling variation is nearly the same as DiM and far from the oracle scenario. Similarly, in the left panel, LS actually has slightly higher sampling variation than DiM when $\mu_{B1}$ is close to the baseline value of $\mu_{A1}=4.7$. The benefit of latent stratification is greatest when $\mu_{A1}$ and $\mu_{B1}$ have different values. To show this pattern more clearly, Figure \ref{fig:sensitivity_mus} shows the percent reductions in sampling variation $\left(1-\frac{Var(\widehat{\tau}^{\text{LS}})}{Var(\widehat{\tau}^{\text{DiM}})}\right)$ for a range of values for $\mu_{A1}$ and $\mu_{B1}$. The lighter area across the diagonal represents the reduced benefit of latent stratification when $\mu_{A1}$ and $\mu_{B1}$ have similar values. 


Figure \ref{fig:sensitivity_pis} shows the sampling variation of the ATE estimators as the strata proportions $\pi_A$, $\pi_B$ and $\pi_C$ are varied. The left panel shows that the reduction in variance for LS is greatest $\pi_C$ is smaller, which means there is more sample available to identify the mixture. The right panel shows that the reduction is higher when $\pi_A$ is a larger relative to $\pi_B$. Figure \ref{fig:sensitivity_pis2} shows the percent reduction in variance across values of $\pi_C$ and $\frac{\pi_A}{\pi_A + \pi_B}$. The surface is relatively flat; in contrast to $\mu_{A1}$ and $\mu_{A0}$, the values of $\pi_A$, $\pi_B$ and $\pi_C$ have less of an effect on the benefits from stratification. 

These simulation results are consistent with our expectations from theory. In general, post-stratification results in lower sampling variance when the individual treatment effects are similar within strata and different between strata \citep{MiratrixSekhonYu2013}. This condition holds by design of the latent strata. In Strata $C$, the individual treatment effects are zero, in $B$ they are all quite large because the treatment acts on the extensive margin, and in $A$, where the treatment acts on the intensive margin, they are somewhere in-between. The ``cost" of latent stratification is that we have to estimate the average effects for each stratum, and the simulation confirms that the model is less well-identified when $\sigma$ is large or when $\mu_{A1}$ and $\mu_{B1}$ have similar values. Under these conditions, the ATE is less precisely estimated relative to post-stratification under the oracle scenario.  When conditions are less than ideal, the LS estimate has about the same precision as DiM. When conditions are ideal for estimating the model, LS has nearly the same level of sampling variation as if the latent strata were observed. 

\section{Discussion \& conclusion}
The recent interest in advertising experiments to estimate incrementality has exposed many challenges that marketers face when fielding and analyzing experiments. Because advertising effects are often small and because consumer response is noisy, precisely estimating these effects is hard even with large samples \citep{LewisRao2015, azevedo2019, berman2021false}. This paper lays out a new approach to analyzing experiments that substantially increases the precision, without requiring any additional covariates. This approach exploits the fact that outcomes in many user-level marketing experiments are positive values with a large number of observed zeros such as quantity purchased, revenue, time-on-site, quantity consumed, etc., and analyzes the data with a model that is well-suited to this structure. This procedure can be used after the experiment is fielded and only requires data on the treatment assignment and outcome. It does not require a special experimental design or any pre-randomization covariates. 

To obtain the benefits from latent stratification, one can apply the following procedure:
\begin{enumerate}
\item Consider whether the key model assumptions are consistent with the setting. The assumptions are (1) there are no customers who would purchase under control, but not under treatment, (2) purchase amounts for the $A$ stratum follow a different distribution than the $B$ stratum, (3) the outcome distributions are Normal with a common variance. If these assumptions are not reasonable, then the LS model is inappropriate or needs modification.
\item Estimate the size of the strata based on how many customers purchase ($Y>0$) under treatment and control (see Section \ref{sec:model}). If one of the strata is very small (e.g. $< 0.1\%$) then latent stratification is inappropriate. 
\item Estimate the LS model and the overall ATE. In our application we estimated the model by MLE and computed the standard error of the ATE using the delta method. 
\item Any issues with convergence of the MLE suggest a lack of identification of the mixture model. This may happen if (1) $\mu_{A1}$ and $\mu_{B1}$ have similar values, and (2) $\sigma$ is large. If any of these problems arise, then latent stratification is inappropriate.

\item Compute the IOS misspecification test. If the specification is rejected, then latent stratification may not be appropriate. 
\item Use cross-validation to compute the prediction MSE and out-of-sample profit of DiM and LS. If DiM outperforms LS out of sample, then LS might not yield a benefit.
\item If there are pre-randomization covariates available that are plausibly related to the individual level treatment effect, stratify on these using a regression adjustment, causal forest or MLRATE. If the sampling error of this approach is better, then consider using it instead of LS.  
\end{enumerate}

Latent stratification is not without limitations. First, latent stratification is only useful for experiments with a continuous response that has a large number of observed zeros. This structure is typical of customer-level sales and could be applied to time-on-site data where there are typically a large proportion of sessions with zero time-on-site (i.e. bounces). However, latent stratification is not appropriate for experiments where the outcome is binary. Second, we make distributional assumptions in the estimation procedure. A non-parametric approach similar to \cite{ding2011} can be applied, but will require using covariates and additional assumptions. Third, the model assumes that there are no customers who would purchase only when they are \emph{not} exposed to the advertising. This assumption improves the identification of the latent stratification model and is consistent with other models of advertising response, e.g. the ad-stock model. A violation of any of these assumptions makes the latent stratification model inappropriate for the data and may result in bias in the LS ATE.

A reliable reduction in the variance of the ATE may be useful for firms in several ways. As we illustrated in Section \ref{sec:decision}, reduced sampling variation directly results in fewer errors and more profit for our example firm deciding whether to discontinue using a marketing channel. Given how noisy incrementality estimates typically are \citep{LewisRao2015, JohnsonLewisReiley2017}, we would likely find the same for other advertisers.  Alternatively, a smaller variance of the ATE translates to a smaller sample size required to detect a positive effect in experiments. When designing experiments the required sample size is linear in the variance of the ATE, and hence a 30\% reduction in variance translates to a 30\% reduction in required sample size. Beyond the benefit of lowering the variance of the ATE, our method also lets firms separate the effects advertising on the intensive and extensive margins. When $\pi_B \mu_{B1}$ is large, the advertising is convincing people to buy, while when $\pi_A(\mu_{A1}-\mu_{A0})$ is large, the advertising is encouraging those who would have bought anyway to buy more. Marketers can test different ad creatives to learn how those creatives affect the intensive and extensive margins.

Other approaches for variance reduction use covariates. Examples include regression adjustment \citep{lin2013}, post-stratification \citep{MiratrixSekhonYu2013}, CUPED \citep{deng2013} and ML methods like causal forests \citep{tibshirani2018package} and MLRATE \citep{guo2021machine}. For these approaches to be successful, the analyst needs access to pre-randomization covariates that are correlated with the potential outcomes observed in the experiments. In our application, we had access to CRM features like recency and frequency, but stratifying on these observed covariates did not reduce sampling variance more latent stratification.  As the online advertising environment marches towards increased data security and consumer privacy, collecting and storing this data also creates potential liability for the firm. In contrast, latent stratification can achieve significant variance reduction  without covariates. While we didn't consider it here, a promising direction is to incorporate covariates into the LS model, either as correlates of the strata membership or as response-shifters. This could reduce reliance on parametric assumptions \citep[cf.][]{ding2011} and further reduce the sampling variance of the ATE. This might also allow us to extend the model to targeting decisions \citep[e.g.][]{simester2022sample} 

The method is not limited to holdout experiments. It also applies to experiments where the treatments are different levels of advertising, under the assumption that an increase in advertising does not decrease sales. Latent stratification can also be applied to pricing experiments. With downward sloping demand, we can assume that a higher price will elicit fewer sales. This means that latent stratification can be applied to experiments on price discounts or price increases, where the higher price is the ``control", and the lower price is the ``treatment." 

Latent stratification could be extended to other settings. Although we focused on the static analysis of experiments that have already been fielded, a Bayesian implementation of the latent stratification model can be readily adapted to dynamic inference in experiments, which is becoming the norm for many A/B testing platforms (e.g. sequential testing, online Bayesian inference, reinforcement learning). This would provide insights about the size and response for each stratum as the experiment is running. We believe such applications hold potential for future work that will combine latent stratification with dynamic experimental designs.

\section{Funding and competing interests}
This work was supported by an Adobe Digital Experience Research Award.
The authors have no other competing interests to report.

\setlength{\bibsep}{0.5pt plus 0.3ex}

\bibliography{references}

\appendix
\setcounter{table}{0}
\renewcommand{\thetable}{\Alph{section}.\arabic{table}}
\renewcommand{\thefigure}{\Alph{section}.\arabic{figure}}

\section{Other applications of principal stratification}
\label{sec:principal_stratification}
Latent stratification is a novel adaptation of principal stratification \citep{FrangakisRubin2002} that is well-suited to advertising experiments where many outcomes are zero. In principal stratification for treatment non-compliance \citep{ImbensRubin1997}, the strata represent potential outcomes for treatment compliance. That is, the strata are defined by whether the treatment each unit actually received ($Z'$) match the randomly assigned treatment ($Z$). The objective is to understand the effect of the applied treatment, which \citeauthor{ImbensRubin1997} dub the complier average causal effect (CACE). This is distinct from latent stratification which does not address treatment non-compliance. See Table \ref{tab:ps_applications}.

Latent stratification is more similar to truncation-by-death \citep{ZhangRubinMealli2009}, where the outcome is undefined for people who have died, e.g., heart rate for an individual who had died before the study endpoint. The strata are defined by whether or not the outcome $Y$ is observed under treatment and control. The objective is to estimate the treatment effect for the $A$ stratum.  Treatment effects for other strata and the overall ATE are undefined \citep{ZhangRubinMealli2009}. 

Latent stratification is different than these other applications because it stratifies based on whether the customer makes a purchase. Our first insight is that not-purchasing has a defined value of zero and we can compute a stratified overall ATE as a weighted average of the strata ATEs. Other applications of principal stratification do not estimate an overall ATE. Our second insight is that the overall ATE has lower sampling variance than the standard difference-in-means. The overall ATE is computed as a weighted average of the observed strata. Computing treatment effects for observed strata and then averaging them together is called post-stratification and can reduce sampling variance \citep{MiratrixSekhonYu2013}. The key difference between our approach and standard post-stratification is that the strata are latent: thus the name ``Latent Stratification".  

\begin{table}[!htb]
\centering
\caption{Key differences between applications of principal stratification}
\label{tab:ps_applications}
\begin{tabular}{lll}
Application & Strata defined by & Goal is to infer:  \\
& potential outcomes for: & \\
\hline
Treatment Non-compliance & $Z' = Z$ & ATE for stratum $B$ (``CACE") \\
Truncation by Death & $Y = \text{NA}$ & ATE for stratum $A$ \\
Latent Stratification & $Y=0$ & Overall ATE \\
\end{tabular}
\end{table}

\section{Distribution of log-sales for the application in Section \ref{sec:applications}}
\setcounter{table}{0}
\setcounter{figure}{0}
\label{sec:app1_hists}

\linespread{0.1}
\begin{figure}[!ht] 
\centering
\includegraphics[width=0.24\textwidth]{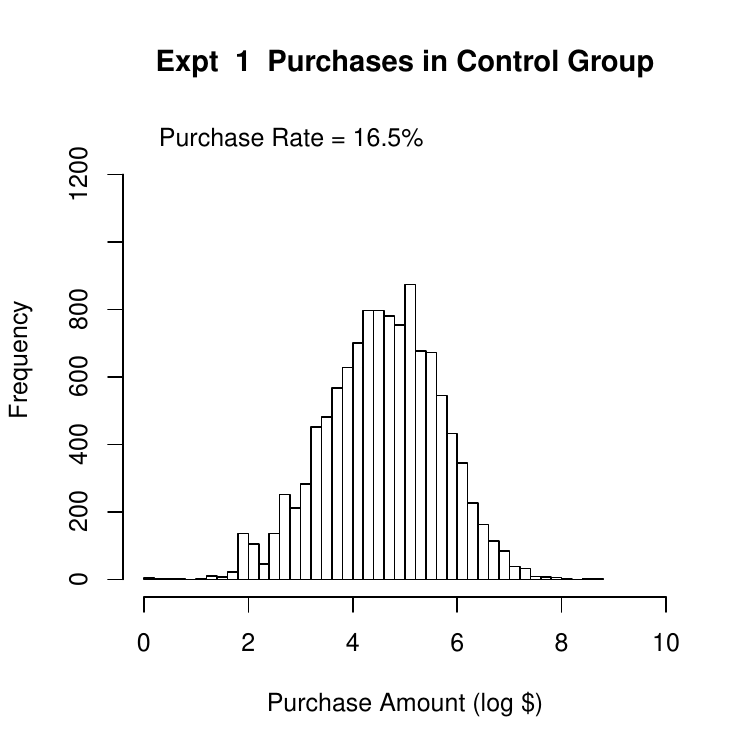}
\includegraphics[width=0.24\textwidth]{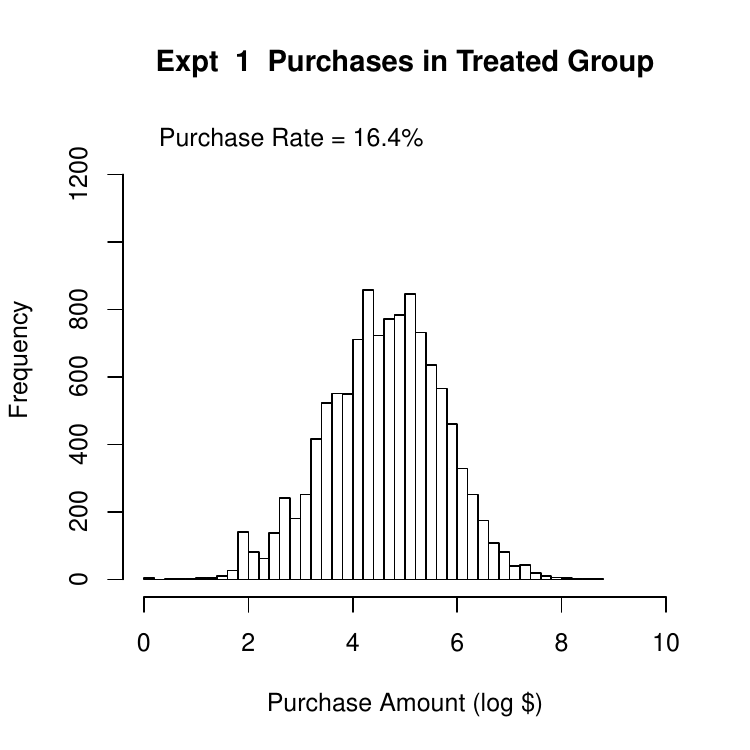}
\includegraphics[width=0.24\textwidth]{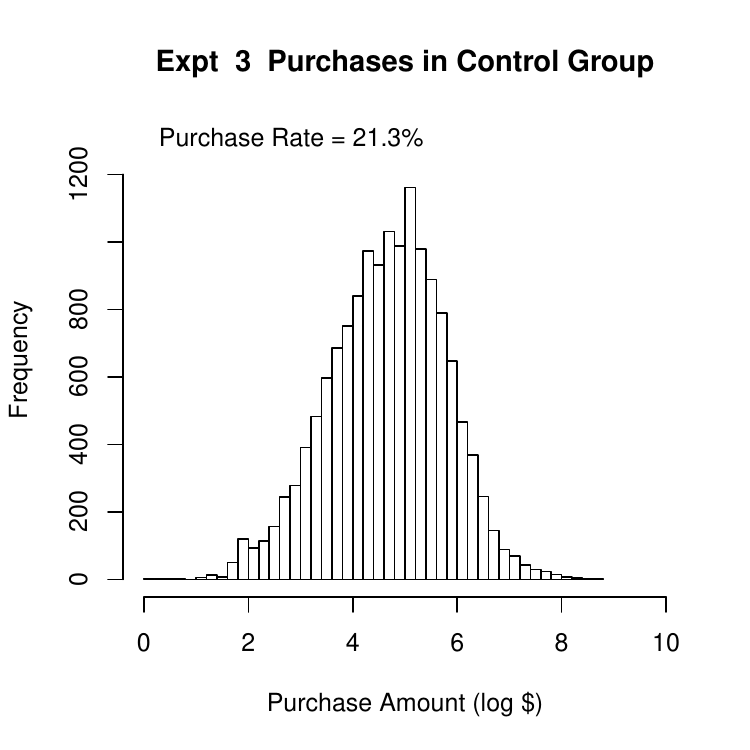}
\includegraphics[width=0.24\textwidth]{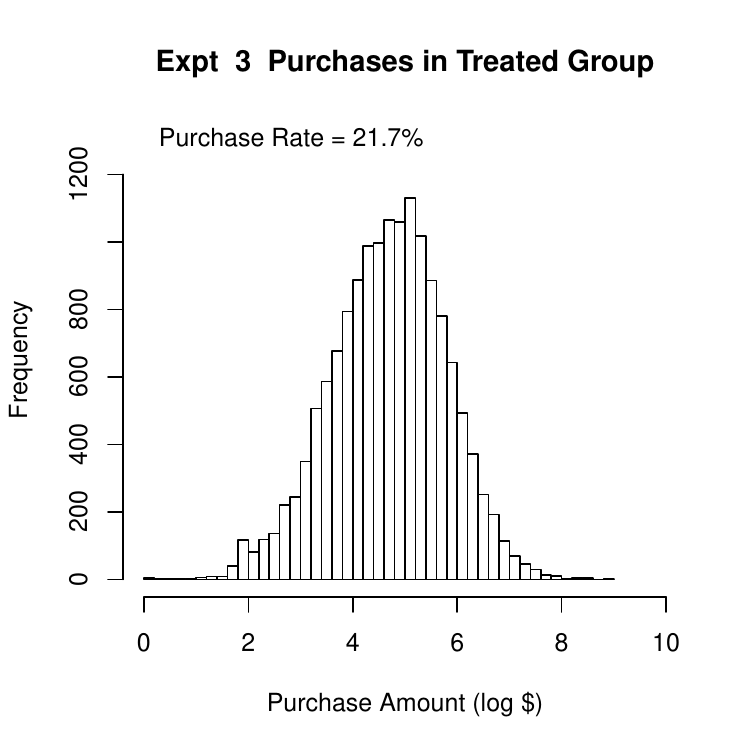}\\
\includegraphics[width=0.24\textwidth]{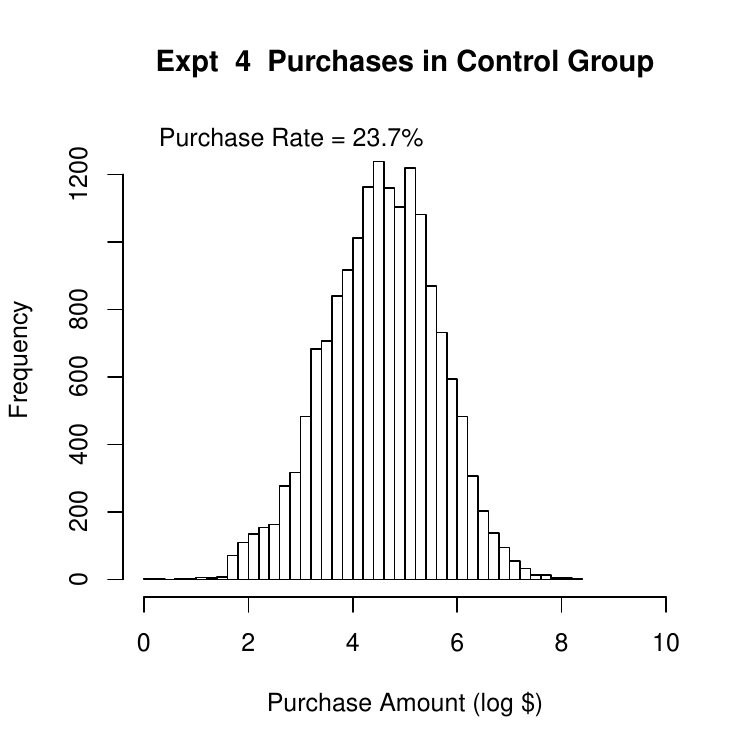}
\includegraphics[width=0.24\textwidth]{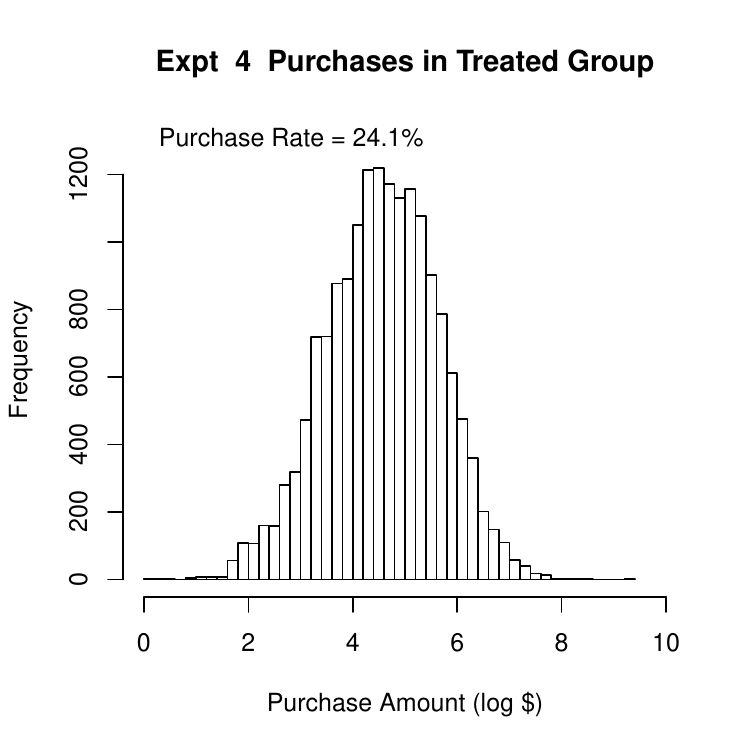}
\includegraphics[width=0.24\textwidth]{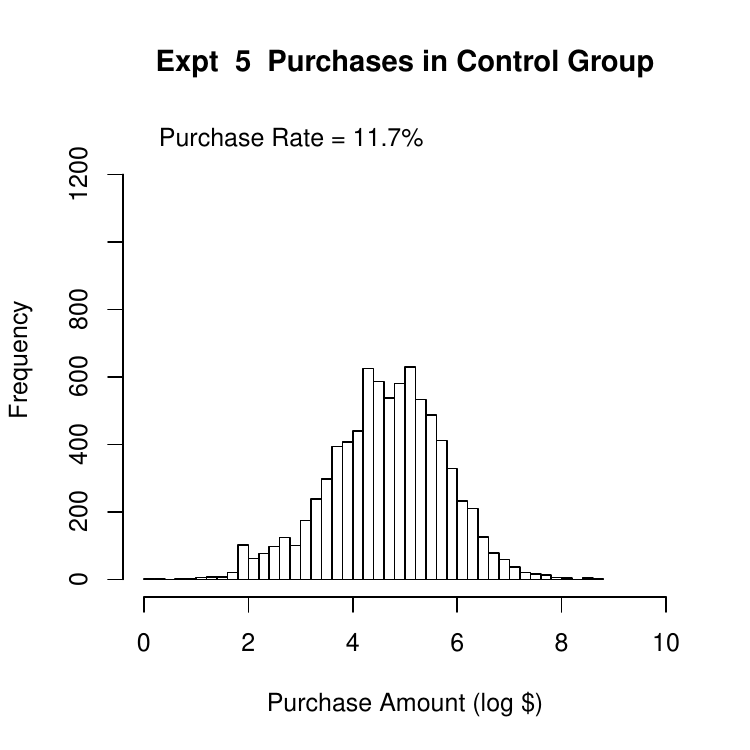}
\includegraphics[width=0.24\textwidth]{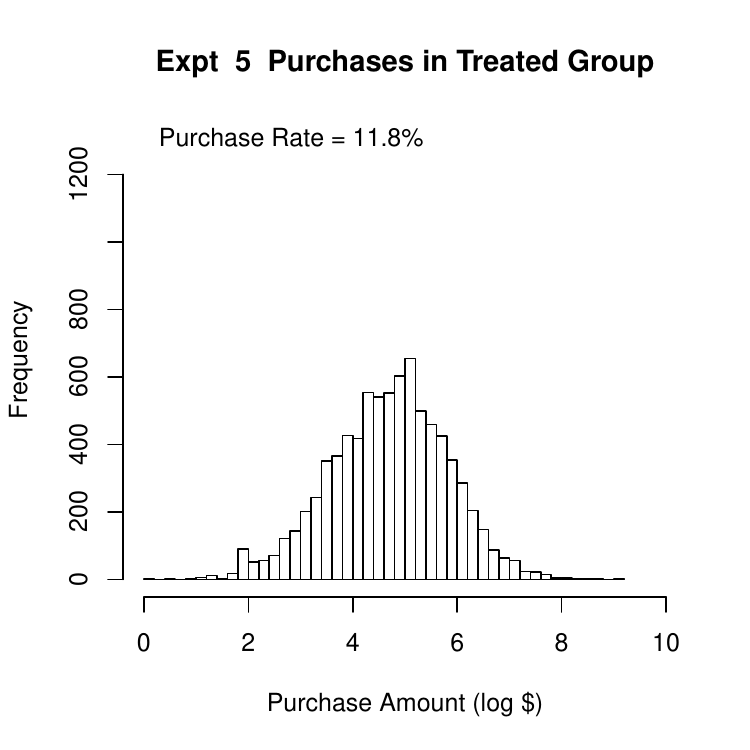}\\
\caption{Distribution of log-purchase amounts for the application in Section \ref{sec:applications}}
\label{fig:app1_hists}
\end{figure}
\linespread{1.5}

\section{Benefit of stratification with known strata memberships (oracle scenario)}
\label{app:benefit}
We assume that the true data generating process is the one described by Equations \eqref{eq:ll_treat_purch}, \eqref{eq:ll_ctrl_purch} and \eqref{eq:ll_zeros}, with the main difference being that the stratum membership of each individual is observed, and derive the variances of $\widehat{\tau}^{\text{DiM}}$ and $\widehat{\tau}^{\text{LS}}$.

The difference-in-means estimator 
$\widehat{\tau}^{\text{DiM}}= \frac{\sum Y_i Z_i}{n_1} - \frac{\sum Y_i (1-Z_i)}{n_0}$
has a population variance of \citep[][Eq. 6.17]{ImbensRubin2015}:
$$Var(\widehat{\tau}^{\text{DiM}}) = \frac{\sigma_0^2}{n_0}+\frac{\sigma_1^2}{n_1}$$ where $\sigma_0^2$ is the population variance of $Y_i(0)$ and $\sigma_1^2$ is the population variance of $Y_i(1)$.

Under the LS model, we derive $\sigma_1^2$ and $\sigma_0^2$ from the mixture components as follows:
\begin{align*}
    \sigma_1^2 & = \pi_A (\sigma^2+\mu_{A1}^2) + \pi_B(\sigma^2+\mu_{B1}^2) -\mu_1^2\\
    &  = \pi_A (\sigma^2+\mu_{A1}^2) + \pi_B(\sigma^2+\mu_{B1}^2) -(\pi_A \mu_{A1} + \pi_B \mu_{B1})^2.
\end{align*}
The first equation comes from the properties of mixture distributions,\footnote{See \url{https://en.wikipedia.org/wiki/Mixture_distribution}, section on Moments, Accessed December 8, 2021.} and the second from plugging in $\mu_1 = \pi_A \mu_{A1} + \pi_B \mu_{B1}$.
Similarly, 
\begin{equation*} 
\sigma_0^2 = \pi_A (\sigma^2+\mu_{A0}^2)-\mu_0^2 = \pi_A (\sigma^2+\mu_{A0}^2) -(\pi_A \mu_{A0})^2.
\end{equation*}

Summarizing, the expression for the variance is:
\begin{equation}
    Var(\widehat{\tau}^{\text{DiM}}) = \frac{\pi_A (\sigma^2+\mu_{A0}^2) -(\pi_A \mu_{A0})^2}{n_0}+\frac{\pi_A (\sigma^2+\mu_{A1}^2) + \pi_B(\sigma^2+\mu_{B1}^2) -(\pi_A \mu_{A1} + \pi_B \mu_{B1})^2}{n_1}
\end{equation}

For the oracle model, the log-likelihood of the data equals:
\begin{align}
    \ell\ell & = \sum_i \left( X_{iA} \log(\pi_A)+ X_{iA} Z_i\log(f_{A1}(Y_i)) + \right.\\
    & + X_{iB} \log(\pi_B) + X_{iB} Z_i \log(f_{B1}(Y_i))+ X_{iC} \log(1-\pi_A-\pi_B)\\
    &\left. + X_{iA} (1-Z_i) \log(f_{A0}(Y_i))\right)
\end{align}
where $f_{A1}(y)$ is the pdf of $\mathcal{N}(\mu_{A1},\sigma^2)$, $f_{A0}(y)$ of $\mathcal{N}(\mu_{A0},\sigma^2)$ and $f_{B1}(y)$ of $\mathcal{N}(\mu_{B1},\sigma_{B1}^2)$, and where $\pi_C$ was replaced by $1-\pi_A-\pi_B$.
Using the log-likelihood we can compute the standard error of the ATE using the inverse of the Fisher Information matrix and the delta method, as follows.

The gradient of the log-likelihood is:
\begin{align}
    \frac{\partial \ell \ell}{\partial\pi_A} & = \sum_i \frac{X_{iA}}{\pi_A} - \sum_i \frac{X_{iC}}{1-\pi_A-\pi_B}\nonumber\\
    \frac{\partial \ell \ell}{\partial\pi_B} & = \sum_i \frac{X_{iB}}{\pi_B} - \sum_i \frac{X_{iC}}{1-\pi_A-\pi_B}\nonumber\\
    \frac{\partial \ell \ell}{\partial\mu_{A1}} & = \sum_i X_{iA}Z_i\frac{Y_i-\mu_{A1}}{\sigma^2}\\
    \frac{\partial \ell \ell}{\partial\mu_{A0}} & = \sum_i X_{iA}(1-Z_i)\frac{Y_i-\mu_{A0}}{\sigma^2}\nonumber\\
    \frac{\partial \ell \ell}{\partial\mu_{B1}} & = \sum_i X_{iB}Z_i\frac{Y_i-\mu_{B1}}{\sigma_{B1}^2}\nonumber\\
    \frac{\partial \ell \ell}{\partial\sigma} & = \sum_i X_{iA}Z_i\frac{(Y_i-\mu_{A1})^2-\sigma^2}{\sigma^3}+\sum_i X_{iA}(1-Z_i)\frac{(Y_i-\mu_{A0})^2-\sigma^2}{\sigma^3}+\sum_i X_{iB}Z_i\frac{(Y_i-\mu_{B1})^2-\sigma^2}{\sigma^3}\nonumber
\end{align}
Denote $\theta = (\pi_A, \pi_B, \mu_{A1}, \mu_{A0}, \mu_{B1},\sigma)^T$, then the Fisher information matrix for one observation $(X_{iA}, X_{iB}, Z_i, Y_i)$ is:
\begin{align}
    & I(\theta) = -\mathbf{E}\left[\frac{\partial^2 LL}{\partial \theta \partial \theta'}\right]\nonumber\\
    & =
    \left(
\begin{array}{cccccc}
 \frac{1}{\pi_A}-\frac{1}{\pi_A+\pi_B-1} & -\frac{1}{\pi_A+\pi_B-1} & 0 & 0 & 0 & 0 \\
 -\frac{1}{\pi_A+\pi_B-1} & \frac{1}{\pi_B}-\frac{1}{\pi_A+\pi_B-1} & 0 & 0 & 0 & 0 \\
 0 & 0 & \frac{p \pi_A}{\sigma ^2} & 0 & 0 & 0 \\
 0 & 0 & 0 & -\frac{(p-1) \pi_A}{\sigma ^2} & 0 & 0 \\
 0 & 0 & 0 & 0 & \frac{p \pi_B}{\sigma ^2} & 0 \\
 0 & 0 & 0 & 0 & 0 & \frac{2 (p \pi_B+\pi_A)}{\sigma ^2} \\
\end{array}
\right)
\end{align}
where $p=Pr(Z_i=1)$.
The Variance-Covariance matrix of $\theta$ is:
\begin{align}
    &V=\left(n I(\theta)\right)^{-1}\nonumber\\
    &=\left(
\begin{array}{cccccc}
 \frac{(1-\pi_A) \pi_A}{n} & -\frac{\pi_A \pi_B}{n} & 0 & 0 & 0 & 0 \\
 -\frac{\pi_A \pi_B}{n} & \frac{(1-\pi_B) \pi_B}{n} & 0 & 0 & 0 & 0 \\
 0 & 0 & \frac{\sigma ^2}{n_1 \pi_A} & 0 & 0 & 0 \\
 0 & 0 & 0 & \frac{\sigma ^2}{n_0 \pi_A} & 0 & 0 \\
 0 & 0 & 0 & 0 & \frac{\sigma ^2}{n_1 \pi_B} & 0 \\
 0 & 0 & 0 & 0 & 0 & \frac{\sigma ^2}{2 n_1 \pi_B+2 n \pi_A} \\
\end{array}
\right)
\end{align}
The gradient of the LS estimator $\widehat{\tau}^{\text{LS}}=\widehat{\pi}_A (\widehat{\mu}_{A1}-\widehat{\mu}_{A0})+\widehat{\pi}_B\widehat{\mu}_{B1}$ equals $$g(\theta) = \frac{\partial g}{\partial \theta} = (\mu_{A1}-\mu_{A0},\mu_{B1},\pi_A,-\pi_A,\pi_B,0)^T.$$
Then by the Delta method:
\begin{align}
    & Var(\widehat{\tau}^{\text{LS}}) = g^T V g\nonumber\\
    & = \frac{\pi_A \sigma ^2}{n_0}+\frac{\sigma ^2 (\pi_A+\pi_B)}{n_1}+\frac{\mu_{B1} \pi_B (2 \mu_{A0} \pi_A-2 \mu_{A1} \pi_A+\mu_{B1})-(\pi_A-1) \pi_A (\mu_{A0}-\mu_{A1})^2-\mu_{B1}^2 \pi_B^2}{n}\nonumber\\
    & = \frac{\pi_A \sigma ^2}{n_0}+\frac{\sigma ^2 (\pi_A+\pi_B)}{n_1}+\frac{\pi_B\mu_{B1}^2+\pi_A (\mu_{A1}-\mu_{A0})^2 - \left(\pi_B \mu_{B1} + \pi_A(\mu_{A1}-\mu_{A0})\right)^2}{n}
\end{align}
Summing up the terms together we get:
\begin{align}
    Var(\widehat{\tau}^{\text{LS}}) & = \pi_A\left(\frac{\sigma^2}{n_{1}}+\frac{\sigma^2}{n_{0}}\right)+(\mu_{A1}-\mu_{A0})^2 \frac{\pi_A(1-\pi_A)}{n}+\nonumber\\& + \pi_B  \frac{\sigma^2}{n_1}+\mu_{B1}^2 \frac{\pi_B (1-\pi_B) }{n}-2\frac{\pi_A \pi_B}{n}(\mu_{A1}-\mu_{A0})\mu_{B1} \label{eq:var_ls_delta}
\end{align}

Finally, the difference between $Var(\widehat{\tau}^{\text{DiM}})$ and $Var(\widehat{\tau}^{\text{LS}})$ equals:
\begin{align}
    Var(\widehat{\tau}^{\text{DiM}})-Var(\widehat{\tau}^{\text{LS}}) & =  \frac{\pi_A \mu_{A0}^2 -(\pi_A \mu_{A0})^2}{n_0}+\frac{\pi_A \mu_{A1}^2 + \pi_B\mu_{B1}^2 -(\pi_A \mu_{A1} + \pi_B \mu_{B1})^2}{n_1} \nonumber \\
        & -\left(\frac{\pi_A(\mu_{A1}-\mu_{A0})^2 +\pi_B\mu_{B1}^2 -\left(\pi_A (\mu_{A1}-\mu_{A0})+\pi_B \mu_{B1}\right)^2}{n}\right)
\end{align}

We can verify that the difference is positive using the following Wolfram Mathematica code:
\begin{verbatim}
diff = (piA*muA0^2 - (piA*muA0)^2)/n0 +
(piA*muA1^2 + piB*muB1^2 - (piA*muA1 + piB*muB1)^2)/n1 -
(piA*(muA1 - muA0)^2 + piB*muB1^2 - (piA*(muA1 - muA0) +piB*muB1)^2)/n

Reduce[{diff > 0, piA > 0, piB > 0, piA + piB < 1,
muA0 > 0, muB1 > 0, muA1 > 0, n1 + n0 == n, n > 0}]
\end{verbatim}

\section{Two-strata estimator}
\label{app:two-strata}
In this section we show that separating consumers into stratum $C$ consumers, and combining the customers in the $A$ and $B$ strata into a single $AB$ stratum, does not yield any variance reduction of the ATE.
We divide the consumers into two strata: stratum $C$ of consumers who don't buy regardless of treatment with size $\pi_C$, and another stratum of potential buyers with size $1-\pi_C$, in which consumers have outcomes with mean $\mu_{AB1}>0$ under treatment and $\mu_{AB0}>0$ under control.

We can estimate the size of the $C$ stratum by counting the number of non-buyers in the treatment group as follows: $\widehat{\pi}_C = \frac{\sum_i \mathbb{I}(Y_i=0) Z_i}{n_1}$.

Using the estimated size of $\pi_C$ we can estimate the ATE in the non-$C$ stratum:
\begin{align}
\widehat{\mu}_{AB1} &= \frac{\sum_i Z_i Y_i}{(1-\widehat{\pi}_C) n_1}\\
\widehat{\mu}_{AB0} &= \frac{\sum_i (1-Z_i) Y_i}{(1-\widehat{\pi}_C) n_0}
\end{align}

The stratified estimator of the ATE equals:
\begin{align}
    \widehat{\tau}^{\text{2S}} & = (1-\widehat{\pi}_C) (\widehat{\mu}_{AB1}-\widehat{\mu}_{AB0})+\widehat{\pi}_C\cdot 0\\
    &= \frac{\sum_i Z_i Y_i}{n_1}-\frac{\sum_i (1-Z_i) Y_i}{n_0} = \widehat{\tau}^{\text{DiM}}
\end{align}
The last equation follows from plugging-in $\widehat{\mu}_{AB1}$ and $\widehat{\mu}_{AB0}$ and noticing that $(1-\widehat{\pi}_C)$ cancels out.
The result is exactly the standard difference-in-means estimator.

\renewcommand{\appendixpagename}{Online Appendix}
\begin{appendices}
\renewcommand{\thesection}{OA.\arabic{section}}
\setcounter{section}{0}
\setcounter{table}{0}
\renewcommand{\thetable}{\thesection.\arabic{table}}
\setcounter{figure}{0}
\renewcommand{\thefigure}{\thesection.\arabic{figure}}
\section{Gradient and Hessian R code}
\label{app:r-code}

The analytical expressions for the observed gradient and Hessian of the latent stratification model can be easily derived using software such as Mathematica by differentiating the log-likelihood directly. Because the resulting expressions are quite long and complex, we include R code for computing these values in the functions \texttt{gr\_ll\_ls()} and \texttt{hes\_ll\_ls()} for the gradient and the Hessian, respectively.

{
\linespread{1.1}
\footnotesize
\begin{verbatim}
# Compute the gradient of the log-likelihood for the latent stratification model. 
#
# par is the vector c(piA, piB, muA1, muA0, muB1, sigma).
# dt is a data frame containing cols y (outcome),
# x (non-zero outcome indicator), z (treatment indicator)
# and xz (TRUE if observation is both treated and has non-zero y). 
#
# Returns the gradient vector with a size of length(par)=6 items.  
#
gr_ll_ls <- function(par, dt) {
  piA <- par[1] 
  piB <- par[2]
  piC <- 1 - piA - piB
  muA1 <- par[3]
  muA0 <- par[4]
  muB1 <- par[5]
  sigma <- par[6]
  
  if (piA < 0 | piB < 0 | piC < 0 | sigma < 0) 
    stop("Error in gr_ll_ls(): piA, piB, piC or sigma < 0")
  
  y <- dt$y
  z <- dt$z
  x <- dt$x
  xz <- dt$xz
  
  y_xz <- y[xz]
  
  # compute Normal density for positive, treated (xz) observations
  fA1 <- dnorm(y_xz, muA1, sigma) # only compute for "treated, purchase" group
  fB1 <- dnorm(y_xz, muB1, sigma) 
  if (sum(is.infinite(c(fA1, fB1)))) 
    warning("Numeric overrun in Normal density calculation in gr_ll_ls()")
  
  # partial derivatives
  dpiA <- sum( fA1/(piA*fA1 + piB*fB1) ) - sum( (1-x)*z )/(1-piA-piB) + 
    sum(x*(1-z))/piA - sum((1-x)*(1-z))/(1-piA)
  dpiB <- sum( fB1/(piA*fA1 + piB*fB1) ) - sum( (1-x)*z )/(1-piA-piB) 
  dmuA1 <- sum( (piA*fA1/(piA*fA1 + piB*fB1))*((y_xz-muA1)/sigma^2) )
  dmuA0 <- sum( x*(1-z)*(y-muA0)/sigma^2 )
  dmuB1 <- sum( (piB*fB1/(piA*fA1 + piB*fB1))*((y_xz-muB1)/sigma^2) )
  dsigma <- sum( (piA*fA1*((y_xz-muA1)^2-sigma^2)/sigma^3 + 
    piB*fB1*((y_xz-muB1)^2-sigma^2)/sigma^3 )/(piA*fA1 + piB*fB1) ) + 
    sum( x*(1-z)*((y-muA0)^2-sigma^2)/sigma^3 )
  
  out <- c(piA=dpiA, piB=dpiB, muA1=dmuA1, muA0=dmuA0, muB1=dmuB1, sigma=dsigma)
}

# Computes the Hessian (matrix of second derivatives)
# of the log-likelihood for the latent stratification model.
# dt is a data frame containing cols y (outcome),
# x (non-zero outcome indicator) and z (treatment indicator). 
#
# Returns the a 6x6 matrix of second derivatives.  
#
hes_ll_ls <- function(par, dt) {
  piA <- par[1] 
  piB <- par[2]
  piC <- 1 - piA - piB
  muA1 <- par[3]
  muA0 <- par[4]
  muB1 <- par[5]
  s <- par[6]
  
  y <- dt$y
  z <- dt$z
  x <- dt$x

  if (piA < 0 | piB < 0 | piC < 0 | s < 0) 
    stop("Error in gr_ll_ls(): piA, piB, piC or sigma < 0")
  
  hes <- matrix(NA, nrow=6, ncol=6)
  Q3 <- exp((muB1-y)^2/(2*s^2))
  Q4 <- exp((muA1-y)^2/(2*s^2))
  Q1 <- (piA*Q3 + piB*Q4)^2 # (piA*exp((muB1-y)^2/(2*s^2))+piB*exp((muA1-y)^2/(2*s^2)))^2
  Q2 <- Q3*Q4 # exp( (muA1^2+muB1^2-2*y*(muA1+muB1)+2*y^2) / (2*s^2) )
  hes[1,1] <- -sum((1-x)*(1-z) / (piA-1)^2)-
    sum((1-x)*z/(piA + piB - 1)^2)-
    sum(x*(1-z)/piA^2)-
    sum(x*z*Q3^2/Q1)
  hes[1,2] <- hes[2,1] <- -sum((1-x)*z/(piA+piB-1)^2) - sum(x*z*Q2/Q1)
  hes[1,3] <- hes[3,1] <- -sum(x*z*piB*(muA1-y)*Q2/(Q1*s^2))
  hes[1,4] <- hes[4,1] <- 0
  hes[1,5] <- hes[5,1] <- sum(x*z*piB*(muB1-y)*Q2/(Q1*s^2))
  hes[1,6] <- hes[6,1] <- sum(x*z*piB*(muA1-muB1)*(muA1+muB1-2*y)*Q2/(Q1*s^3))
  hes[2,2] <- -sum((1-x)*z/(piA+piB-1)^2) - sum(x*z*Q4^2/Q1)
  hes[2,3] <- hes[3,2] <- sum(x*z*piA*(muA1-y)*Q2/(Q1*s^2))
  hes[2,4] <- hes[4,2] <- 0
  hes[2,5] <- hes[5,2] <- -sum(x*z*piA*(muB1-y)*Q2/(Q1*s^2))
  hes[2,6] <- hes[6,2] <- -sum(x*z*piA*(muA1-muB1)*(muA1+muB1-2*y)*Q2/(Q1*s^3))
  hes[3,3] <- -sum(x*z*Q3*piA*(Q3*piA*s^2+Q4*piB*(-muA1^2+s^2+2*muA1*y-y^2))/(Q1*s^4))
  hes[3,4] <- hes[4,3] <- 0
  hes[3,5] <- hes[5,3] <- -sum(x*z*piA*piB*(muA1-y)*(muB1-y)*Q2 / (Q1*s^4))
  hes[3,6] <- -sum(x*z*Q3*piA*(y-muA1)*
    (2*Q3*piA*s^2+Q4*piB*(muB1^2-muA1^2+2*s^2+2*muA1*y-2*muB1*y)) / (Q1*s^5))
  hes[6,3] <- hes[3,6]
  hes[4,4] <- -sum(x*(1-z)/s^2)
  hes[4,5] <- hes[5,4] <- 0
  hes[4,6] <- hes[6,4] <- sum(x*(1-z)*2*(muA0-y)/s^3)
  hes[5,5] <- -sum(x*z*piB*Q4*(Q4*piB*s^2 + Q3*piA*(s^2+2*muB1*y-muB1^2-y^2))/(Q1*s^4))
  hes[5,6] <- -sum(x*z*piB*(y-muB1)*Q4*(2*Q4*piB*s^2+Q3*piA*(muA1^2-muB1^2+2*s^2-2*muA1*y+2*muB1*y))/(Q1*s^5))
  hes[6,5] <- hes[5,6] 
  hes[6,6] <- sum(x*(1-z)*(s^2-3*(muA0-y)^2)/(s^4))+
    sum(x*z*(Q3^2*piA^2*s^2*(s^2-3*(muA1-y)^2)+
    Q4^2*piB^2*s^2*(s^2-3*(muB1-y)^2)+
    Q2*piA*piB*(muA1^4 - 2*muA1^2*muB1^2 + muB1^4 - 3*muA1^2*s^2 - 3*muB1^2*s^2 + 
    2*s^4 - 2*(muA1 + muB1)*(2*(muA1 - muB1)^2 - 3*s^2)*y + 
    2*(2*(muA1 - muB1)^2 - 3*s^2)*y^2))/(Q1*s^6))

  hes
}
\end{verbatim}
}

\section{Identification of two-component Normal mixture models} 
\label{sec:mixture_identification}
A general two-component Normal mixture model can be poorly-identified \citep{ho2022weak}. However, as we show in this section, the LS model is much better identified because data observed in the control group serves to constrain the mixture model for the treated customers who buy. As we discuss in Section \ref{sec:ls}, the mixing proportion between the two components ($\pi_A/(\pi_A + \pi_B)$) is well-identified by the difference in purchase rates in treatment and control. Further, in our application, we assume that the variance of the Normal components is the same for A0, A1 and B0, and that common $\sigma$ is well-identified by the observed variance for customers in control who purchase (A0). This additional structure in the LS model substantially improves the identification of the mixture. 

To illustrate this, we fit an unconstrained two-component normal mixture using the data for treated customers that buy in Experiment 2. That is, we fit the model with the likelihood
\begin{equation}
\ell(Y_i) = \pi \frac{1}{\sigma_{A1}}\phi\left(\frac{Y_i-\mu_{A1}}{\sigma_{A1}}\right) + (1-\pi) \frac{1}{\sigma_{B1}} \phi\left(\frac{Y_i-\mu_{B1}}{\sigma_{B1}}\right)
\end{equation}
using the data that identifies the mixture model in LS. We then compare that to the LS parameter estimates using the full data set. 

Table \ref{tab:mixture_identification} shows parameter estimates for a general mixture and the LS model along with bootstrapped standard errors. The parameters for the unconstrained mixture model have large standard errors, particularly for the mixing ratio $\pi_A/(\pi_A + \pi_B)$\footnote{The standard mixing ratio in the two-component normal mixture is $\pi_A/(\pi_A + \pi_B)$. We estimate this directly for the mixture model and indirectly for the LS model.} and the parameters of the smaller mixture component $\mu_{B1}$ and $\sigma_{B1}$. By contrast, the corresponding parameters in the LS model have standards errors that are an order of magnitude smaller. The LS model is substantially better identified than the unconstrained mixture model. 

\linespread{1.1}
\begin{table}[!htbp] 
\centering 
\caption{Comparison of parameter estimates for an unconstrained two-component normal mixture model versus the latent stratification model for Experiment 2 with bootstrap standard errors shown in parentheses.} 
\label{tab:mixture_identification} 
\begin{tabular}{@{\extracolsep{5pt}} ccc} 
\toprule 

 & Unconstrained & Latent \\ 
 & Mixture Model &  Stratification \\
\midrule
$\pi_A/(\pi_A+\pi_B)$ & 0.870 & 0.978\\ 
& (0.243) & (0.008) \\ 
[1ex]
$\mu_{A1}$ & 4.860 & 4.688\\ 
& (0.147) & (0.017) \\ 
[1ex]
$\mu_{B1}$ & 3.258 & 2.992\\ 
& (0.691) & (0.439) \\ 
[1ex]
$\sigma_{A1}$ & 0.979 & 1.102 \\ 
 & (0.007) & (0.007) \\ 
[1ex]
$\sigma_{B1}$ & 0.894 &  \\ 
& (0.303) &  \\ 
[1ex]
$\pi_C$ &  & 0.834 \\ 
&  & (0.001) \\ 
[1ex]
$\mu_{A0}$ &  & 4.616 \\ 
&  &  (0.011) \\ 
\midrule
$\ell\ell$ treated, buy & -17371.6 & -17398.2 \\ 
$N$ treated, buy& 11,442 & 11,442 \\ 
[1ex]
$\ell\ell$ all & & -96265.8 \\ 
$N$ all &  & 138,227 \\ 
\hline \\[-1.8ex] 
\end{tabular} \\
\end{table}

\newpage
\section{Alternative benefit of LS: reduced sample sizes}
\label{app:sample-sizes}

An alternative way to realize the benefit of latent stratification is to reduce the size of the total sample needed for an experiment to detect a specific increase in sales. The retailer that provided the data estimated that the cost of sending a catalog to customers was approximately one US dollar. An increase of one dollar on average in Experiment 2 between treatment and control translates to approximately a 5\% increase in sales (and also a 5\% increase in log-sales). Using a DiM analysis would have required approximately 72,000 consumers to detect such an effect, while using LS would have required approximately 38,000 consumers (almost half). 

To compute the required sample size with equal allocation between treatment and control, we use the formulas:
$$n_1 = \left(1+\frac{1}{\kappa}\right)(z_{\alpha/2}+z_{\beta})^2\frac{\sigma^2}{d^2}\hspace{0.5in}n_0 = \kappa n_1$$
where $\alpha=0.05$, $\beta=0.2$, $z_q$ is the upper $q$ quantile of the standard Normal distribution,  $\sigma$ is the data's standard deviation, $d$ is the effect size to be detected and $\kappa$ is the ratio of control to treatment group sample sizes $\kappa=n_0/n_1$. The result is the required sample for each arm (treatment or control).
Suppose we would like to detect a log-sales increase of 5\% vs.~the null hypothesis of no increase in log-sales. In Experiment 2 this increase would translate to an increase from log-sales of 0.747 to 0.784, or about \$1.06 in sales, which would cover the costs of sending the catalog to customers.

Using table \ref{tab:app1_var_red}, we can compute the standard deviation of the data under difference-in-means to be approximately $0.0095\cdot \sqrt{70000/2}=1.777$, while for latent stratification it is $0.0069\cdot \sqrt{70000/2}=1.29$. Plugging into the sample size formula, under difference-in-means we would need approximately 36k consumers per treatment, while under latent stratification we would need approximately 19k consumers per treatment.

Another way to realize the benefit is to use a substantially smaller control group. We can vary the ratio of sample sizes of the control and the treatment groups $\kappa$ to make the control group as small as possible while leaving the total sample size fixed. For Experiment 2 with $n=138,227$, in order to detect the same 5\% increase in sales this implies that $\kappa$ can be as low as 8.1\%, yielding $n_0 = 10,307$ and $n_1 = 127,920$. If the true effect is an average increase in sales of \$1.19 (as it is in the data for Experiment 2), and assuming a catalog cost of \$1, this approach could increase the firm's profit earned during the test by $(68914-10307) \cdot 0.19 = \$11,135$. 

We can also look at the increase in power due to latent stratification given a fixed sample size of 140k consumers, equally allocated between treatment and control. The smaller standard error would have increased the power to detect a 3\% increase in sales from 64\% under difference-in-means to 89\%.

\section{Regression adjustment with observed pre-randomization covariates}
\label{app:regression_adjustment}

\begin{table}[ht]
\scriptsize
\centering
\caption{Treatment effects with post-stratification using pre-rendomization covariates}
\begin{tabular}{lllllll}
\toprule
                                & Experiment 40         & Experiment 41        & \multicolumn{2}{l}{Experiment 42}        & Experiment 43 & Experiment 48 \\
\midrule
Treatment                       & 0.0044                & 0.0252**             & \multicolumn{2}{l}{0.0256**}             & 0.0257**      & 0.0112        \\
                                & (0.0086)              & (0.0087)             & \multicolumn{2}{l}{(0.0097)}             & (0.0099)      & (0.0078)      \\
T                               & 0.5431***             & 0.6148***            & \multicolumn{2}{l}{0.8229***}            & 0.8493***     & 0.4806***     \\
                                & (0.0349)              & (0.0358)             & \multicolumn{2}{l}{(0.0396)}             & (0.0413)      & (0.0320)      \\
D                               & 0.0007*               & 0.0009+              & \multicolumn{2}{l}{0.0005}               & 0.0009+       & 0.0008+       \\
                                & (0.0003)              & (0.0005)             & \multicolumn{2}{l}{(0.0004)}             & (0.0005)      & (0.0004)      \\
I                               & -0.1803***            & -0.0833*             & \multicolumn{2}{l}{0.0162}               & 0.0285        & -0.1234***    \\
                                & (0.0365)              & (0.0397)             & \multicolumn{2}{l}{(0.0411)}             & (0.0422)      & (0.0367)      \\
M                               & 0.0018*               & 0.0014               & \multicolumn{2}{l}{0.0021*}              & 0.0010        & 0.0016*       \\
                                & (0.0007)              & (0.0009)             & \multicolumn{2}{l}{(0.0008)}             & (0.0009)      & (0.0008)      \\
C                               & 0.0019**              & 0.0025**             & \multicolumn{2}{l}{0.0013}               & 0.0020*       & 0.0020*       \\
                                & (0.0007)              & (0.0009)             & \multicolumn{2}{l}{(0.0008)}             & (0.0009)      & (0.0009)      \\
F                               & 3.1164***             & 2.9989***            & \multicolumn{2}{l}{3.4019***}            & 3.3151***     & 2.1056***     \\
                                & (0.0930)              & (0.0948)             & \multicolumn{2}{l}{(0.0945)}             & (0.0883)      & (0.0960)      \\
R                               & 0.0058***             & 0.0050**             & \multicolumn{2}{l}{-0.0052**}            & -0.0114***    & 0.0049**      \\
                                & (0.0017)              & (0.0017)             & \multicolumn{2}{l}{(0.0020)}             & (0.0020)      & (0.0016)      \\
Treatment ×  T         & -0.0069               & 0.0150               & \multicolumn{2}{l}{0.0904}               & 0.0875        & -0.0127       \\
                                & (0.0491)              & (0.0504)             & \multicolumn{2}{l}{(0.0557)}             & (0.0580)      & (0.0448)      \\
Treatment ×  D         & 0.0000                & -0.0009              & \multicolumn{2}{l}{-0.0008}              & -0.0010+      & -0.0002       \\
                                & (0.0005)              & (0.0006)             & \multicolumn{2}{l}{(0.0005)}             & (0.0006)      & (0.0005)      \\
Treatment ×  I         & 0.0035                & -0.0232              & \multicolumn{2}{l}{-0.0167}              & 0.0573        & 0.0219        \\
                                & (0.0509)              & (0.0552)             & \multicolumn{2}{l}{(0.0567)}             & (0.0577)      & (0.0503)      \\
Treatment ×  M         & -0.0004               & 0.0024*              & \multicolumn{2}{l}{0.0023*}              & 0.0022+       & 0.0012        \\
                                & (0.0010)              & (0.0011)             & \multicolumn{2}{l}{(0.0011)}             & (0.0011)      & (0.0010)      \\
Treatment ×  C         & 0.0007                & -0.0022+             & \multicolumn{2}{l}{-0.0016}              & -0.0018       & -0.0008       \\
                                & (0.0010)              & (0.0012)             & \multicolumn{2}{l}{(0.0011)}             & (0.0011)      & (0.0011)      \\
Treatment ×  F         & 0.0060                & 0.0406               & \multicolumn{2}{l}{-0.1193}              & -0.1604       & -0.1184       \\
                                & (0.1256)              & (0.1285)             & \multicolumn{2}{l}{(0.1299)}             & (0.1232)      & (0.1270)      \\
Treatment ×  R         & -0.0016               & 0.0005               & \multicolumn{2}{l}{0.0018}               & -0.0011       & -0.0013       \\
                                & (0.0025)              & (0.0025)             & \multicolumn{2}{l}{(0.0028)}             & (0.0028)      & (0.0023)      \\
Constant                        & -0.1580***            & -0.1753***           & \multicolumn{2}{l}{-0.0179}              & 0.1133***     & -0.1555***    \\
                                & (0.0226)              & (0.0226)             & \multicolumn{2}{l}{(0.0251)}             & (0.0255)      & (0.0207)      \\
                                \midrule
N                               & 138,281                & 138,227               & \multicolumn{2}{l}{138,155}               & 138,138        & 138,039        \\
Adjusted $R^2$                     & 0.182415              & 0.175661             & \multicolumn{2}{l}{0.181214}             & 0.170583      & 0.130446      \\
\midrule
\multicolumn{4}{l}{+ p \textless 0.1, * p \textless 0.05, ** p \textless 0.01, *** p   \textless 0.001} & \multicolumn{3}{l}{}                            \\
\multicolumn{4}{l}{Heteroskedasticity robust standard errors are in   parentheses.}                & \multicolumn{3}{l}{}\\                           
\bottomrule
\end{tabular}
\end{table}

\section{Accuracy of the LS ATE}
\label{sec:accuracyLSATE}
\begin{figure}[!htb]
\centering
\includegraphics[width=0.49\textwidth, page=6]{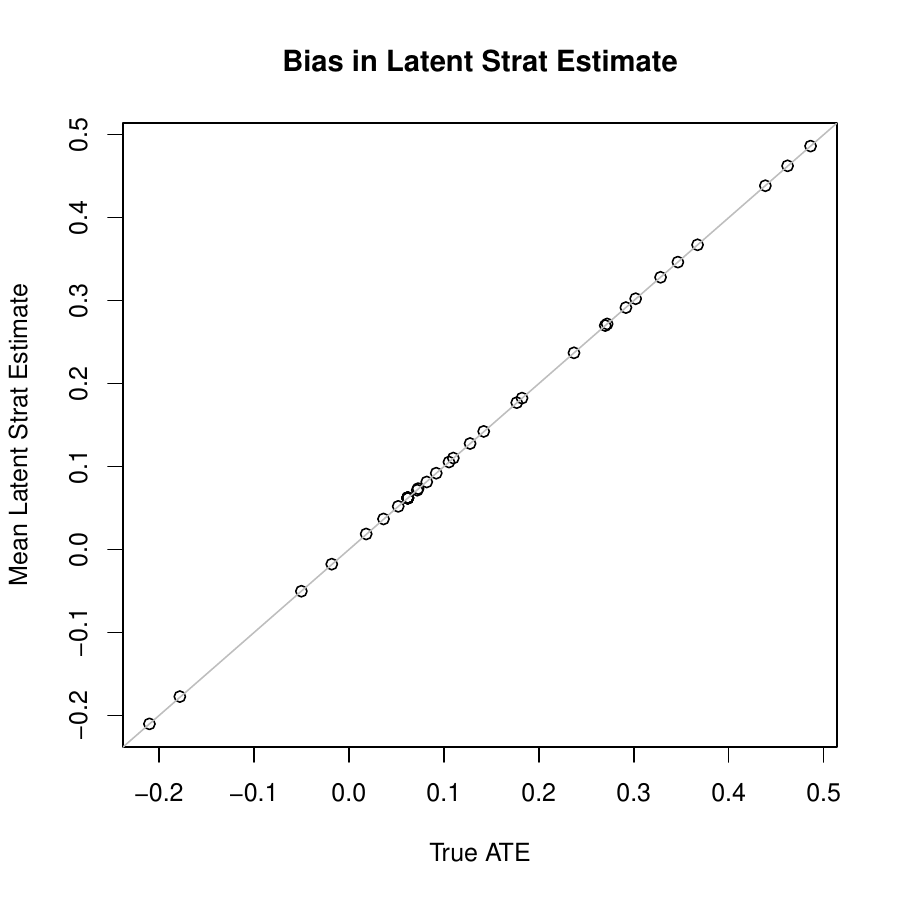}
\includegraphics[width=0.49\textwidth, page=9]{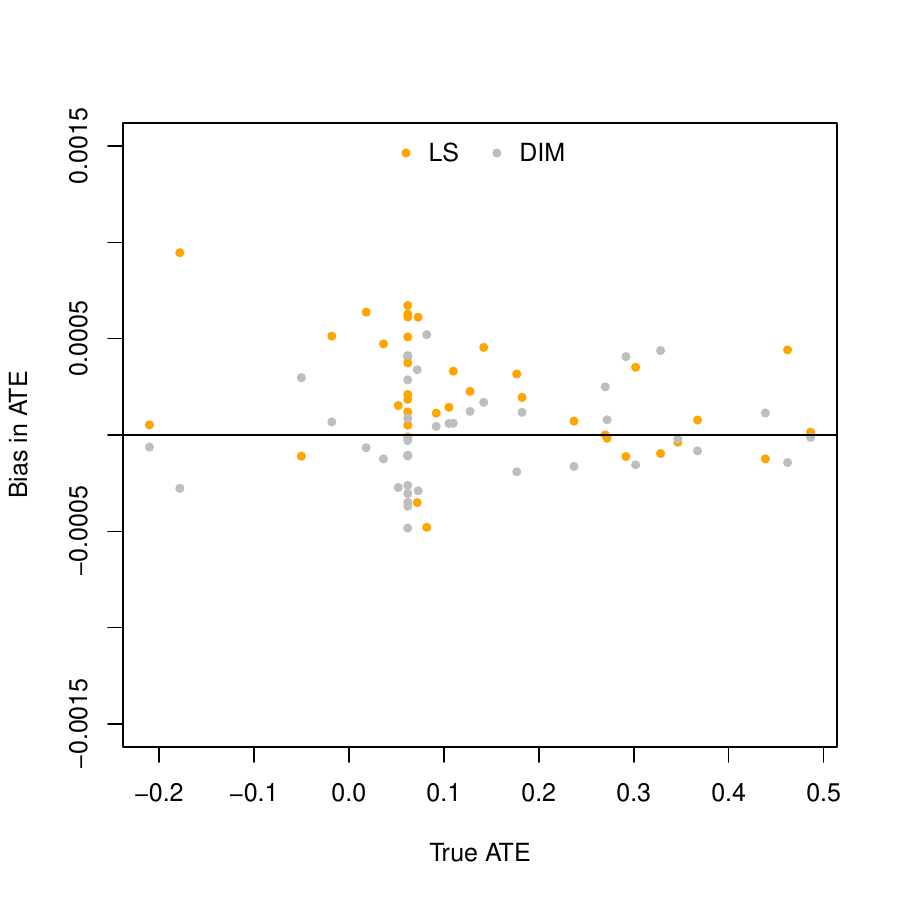}
\caption{Simulation study shows empirical bias in $\widehat{\tau}^{\text{LS}}$ is minimal.}
\label{fig:bias_comparison}
\end{figure}

\textbf{When the model is correctly specified}, the LS ATE is consistent, like any other maximum-likelihood estimate. However, it may be biased in finite samples. In a simulation study, we find that this finite sample bias is minimal with a sample size of 100,000 (50,000 in treatment and 50,000 in control). For each set of parameter values in the simulation, we compute the average estimate of the ATE (averaged over the 2000 simulated data sets) and compare it to the true value to obtain an empirical estimate of the bias of $\widehat{\tau}^{\text{LS}}$. For the base parameter values, the mean estimate is 0.06221 versus a true ATE of 0.06200 (0.34\% empirical bias). The left panel of Figure \ref{fig:bias_comparison} plots the true ATE versus the mean of $\widehat{\tau}^{\textnormal{LS}}$ for the 37 different parameter settings in the simulation study and shows that the bias is minimal across parameter settings. The right panel shows that $\widehat{\tau}^{\text{LS}}$ has modest positive bias relative to $\widehat{\tau}^{\text{DiM}}$. (Recall, DiM is is unbiased in finite samples, so any bias we find there is due to sampling variation in the simulation.) Thus, we conclude that $\widehat{\tau}^{\text{LS}}$ is reasonably close to unbiased for our example application with sample size around 70,000 each in treatment and control. 

\begin{figure}[!htb]
\centering
\includegraphics[width=0.5\textwidth, page=2]{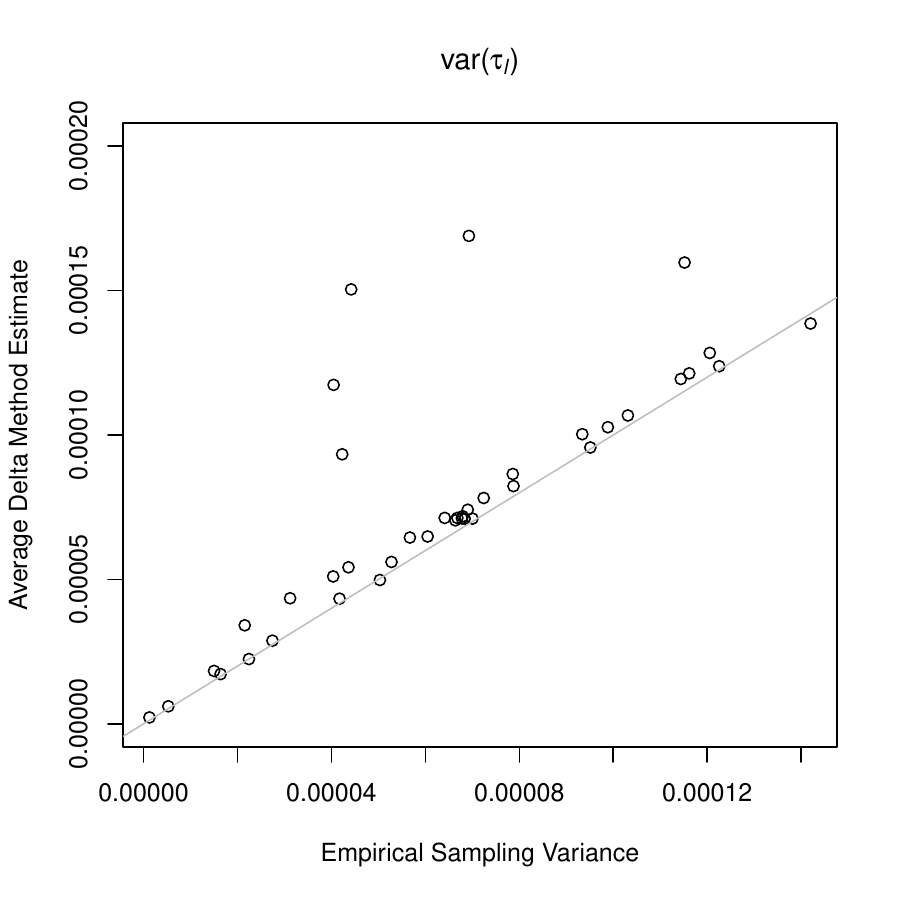}
\caption{Empirical sampling variation of $\widehat{\tau}^\text{LS}$ versus delta method estimate of sampling variation}
\label{fig:delta_method_check}
\end{figure}

To confirm that the delta method produces a reliable estimate of the sampling variation of the ATE ($\widehat{\tau}^{\text{LS}}$), Figure \ref{fig:delta_method_check} plots the empirical sampling variation from the simulation versus the average delta method estimate (across the 2000 simulated data sets). The plot shows that the delta method estimates are typically a little higher than the true sampling variation of the estimator. This suggests that the delta method provides a conservative estimate of the sampling variation in the MLE estimator, conditional on the model being correctly specified. 

\textbf{When the model is misspecified,} the LS ATE may be biased. Thus, it is important for users to consider whether the LS assumptions hold for their data and to use the misspecification test to detect misspecification. (See Section \ref{sec:assumptions}.) In this section, we quantify the bias in $\widehat{\tau}^{LS}$ under three different potential misspecifications: 1) the outcomes are not normally distributed, 2) $\sigma_{B1} \ne \sigma_{A1} = \sigma_{A0}$, and 3) there is a fourth stratum that purchases under control, but not treatment. We also test the power of the IOS test to detect these misspecifications.

We extend the simulations to account for misspecification, by generating a synthetic data set with $N=100,000$ customers from a model similar to LS, with parameters $\pi_A=0.16$, $\pi_B=0.01$, $\pi_C=0.83$, $\mu_{A1}=4.7$, $\mu_{A0}=4.5$ and $\mu_{B1}=3.0$. However, to create misspecification of the outcome distributions we add error terms drawn from a fatter-tailed t distribution with 3, 7 or 10 degrees of freedom or from a skewed shifted-Gamma distribution. The mean of the error terms is fixed to zero with standard deviation 1 (i.e. $\sigma_{A1} = \sigma_{A0} = \sigma_{B1}=1$). We then compute $\widehat{\tau}^{LS}$ and $\widehat{\tau}^{DiM}$ using this data. We also computed the IOS test statistic and p-value using 100 bootstrapped samples from the simulated data set. We repeated this process with 100 simulated data sets, allowing us to estimate the bias of the LS estimator and the IOS test rejection rate at $p < 0.1$. 

The results summarized in Table \ref{tab:misspec_sim1} show that the bias in $\widehat{\tau}^{LS}$ is rather modest. The magnitude of the estimated bias is 0.001 for both $\widehat{\tau}^{DiM}$ and $\widehat{\tau}^{LS}$, suggesting that the bias in $\widehat{\tau}^{LS}$ is within the sampling error of our procedure.\footnote{Since $\widehat{\tau}^{DiM}$ is unbiased, the estimated bias in $\widehat{\tau}^{DiM}$ represents sampling error.} We also find that the IOS test fails to detects these misspecifications. 

\begin{table}[!htbp] \centering 
\caption{Estimated bias in $\widehat{\tau}^{LS}$ when simulated error terms follow t or Gamma distributions.} 
\label{tab:misspec_sim1} 
\begin{tabular}{lcccc} 
\toprule
 Error Distribution & \multicolumn{1}{c}{Bias in $\widehat{\tau}^{LS}$} & \multicolumn{1}{c}{Bias in $\widehat{\tau}^{DiM}$} & \multicolumn{1}{c}{IOS Rejection Rate} & \multicolumn{1}{c}{Mean IOS p-value} \\ 
\midrule
$t(3)$ & -0.001 & 0.001 & 0.10 & 0.482  \\ 
$t(7)$ & 0.001 & 0.001 & 0.12 & 0.463 \\ 
$t(10)$ & 0.001 & 0.001  & 0.08 & 0.520 \\ 
Gamma & 0.000 & 0.001 &  0.10 & 0.537 \\ 
\bottomrule
\end{tabular} 
\end{table} 

We completed a similar simulation where the data was generated from Normal distributions, but with differing values of $\sigma_{B1} \ne \sigma_{A1} = \sigma_{A0} = 1$. Table \ref{tab:misspec_sim2} shows that $\widehat{\tau}^{LS}$ can be substantially biased when $\sigma_{B1}$ is larger or smaller than $\sigma_{A1} = \sigma_{A0} = 1$. The IOS test reliably detects this misspecification when  $\sigma_{B1}$  is smaller than $\sigma_{A1} = \sigma_{A0} = 1$, but not when $\sigma_{B1}$ is larger than $\sigma_{A1} = \sigma_{A0} = 1$.

\begin{table}[!htbp] \centering 
\caption{Estimated bias in $\widehat{\tau}^{LS}$ when $\sigma_{B1} \ne \sigma_{A1} = \sigma_{A0} = 1$.} 
\label{tab:misspec_sim2} 
\begin{tabular}{lrrrr} 
\toprule
 $\sigma_{B1}$ & \multicolumn{1}{c}{Bias in $\widehat{\tau}^{LS}$} & \multicolumn{1}{c}{Bias in $\widehat{\tau}^{DiM}$} & \multicolumn{1}{c}{IOS Rejection Rate} & \multicolumn{1}{c}{Mean IOS p-value} \\ 
\midrule
0.75 & 0.012 & -0.002 & 0.95 & 0.026  \\ 
1.25 & -0.015 & -0.001 & 0.00 & 0.912 \\ 
\bottomrule
\end{tabular} 
\end{table} 

Finally, we repeated this analysis generating data from a model where there are four strata: the three strata in the LS model and a fourth stratum ``D" where $Y(0)>1$ and $Y(1)=0$. We assumed that the response for the control group stratum D is normal with mean $\mu_D = 4$ and standard deviation $\sigma_{D0} = \sigma_{A1} = \sigma_{A0} = \sigma_{B1}$. We varied the size of stratum D from $\pi_D = 0$ to $0.015$, reducing the size of the A and B strata proportionally. As Table \ref{tab:misspec_sim3} shows, this misspecification produces substantial biases in $\widehat{\tau}^{LS}$. This misspecification is readily detected by the IOS test when the D stratum is large ($\pi_D=0.015$), however there is still substantial bias when $\pi_D=0.005$ and this is only detected by the IOS test about half the time. 

\begin{table}[htbp!]
\centering 
\caption{Estimated bias in $\widehat{\tau}^{LS}$ when there is a fourth strata where $Y(0)>1$ and $Y(1)=0$. The size of this group is $\pi_D$.} 
\label{tab:misspec_sim3} 
\begin{tabular}{@{\extracolsep{5pt}} crrrrrrrr} 
\toprule
 $\pi_{D}$ & \multicolumn{1}{c}{Bias in $\widehat{\tau}^{LS}$} & \multicolumn{1}{c}{Bias in $\widehat{\tau}^{DiM}$} & \multicolumn{1}{c}{IOS Rejection Rate} & \multicolumn{1}{c}{Mean IOS p-value} \\ 
\midrule
0.000 & 0.002 & -0.002 & 0.07 & 0.525 \\ 
0.005 & 0.012 & -0.001 & 0.46 & 0.226  \\ 
0.010 & 0.027 & -0.001 & 0.77 & 0.084 \\ 
0.015 & 0.044 & 0.001 & 0.95 & 0.030 \\ 
\hline 
\hline \\[-1.8ex] 
\end{tabular} 
\end{table}

\end{appendices}

\end{document}